%% file: dtw-for-few-shot.tex
\pdfoutput=1
\documentclass[11pt,a4paper]{article}
\usepackage[hyperref]{acl2020}
\usepackage{times}
\usepackage{latexsym}

\usepackage{microtype}

\aclfinalcopy

\usepackage{comment}
\usepackage{url}
\usepackage{textcomp}

\input{extras}

\input{preamble}
\usepackage{csquotes}

\begin{document}

\input{metadata}

\maketitle

\begin{abstract}
\input{abstract}
\end{abstract}

\input{main}

\bibliographystyle{acl_natbib}
\bibliography{bibliography}

\end{document}

%% file: extras.tex
\input{applica-extras}

\input{autozoil-extras}

%% file: applica-extras.tex
\usepackage[T1]{fontenc}
\usepackage[utf8]{inputenc}
\usepackage{graphicx}
\usepackage{booktabs}

\usepackage{footnote}
\makesavenoteenv{tabular}
\makesavenoteenv{table}
\makesavenoteenv{table*}

\usepackage{xurl}

\newcommand\minput[1]{%
  \input{#1}%
  \ifhmode\ifnum\lastnodetype=11 \unskip\fi\fi}

\usepackage{hyperref}
\usepackage{xstring}



%% file: autozoil-extras.tex
\newcommand{\noqa}[1]{}

%% file: preamble.tex
\usepackage{amsmath}

\DeclareMathOperator*{\argmin}{arg\,min}
\usepackage{mathtools}
\usepackage{amsmath,amsfonts,amssymb}
\usepackage{algpseudocode}
\usepackage{nicefrac}
\usepackage{algorithm}
\usepackage[most]{tcolorbox}

\usepackage{tabularx}
\usepackage{multirow}
\usepackage{natbib}
\usepackage{color,soul}
\usepackage{nomencl}
\makenomenclature

%% file: metadata.tex
\title{Dynamic Boundary Time Warping\\for Sub-sequence Matching with Few Examples}

\author{Łukasz Borchmann\thanks{\quad Equal contribution.} \\
  Applica.ai \\ Warsaw, Poland \\
  \small{lukasz.borchmann@applica.ai}\\
  \\\And
  Dawid Jurkiewicz\footnotemark[1] \\
  Applica.ai \\ Warsaw, Poland \\
  \small{dawid.jurkiewicz@applica.ai}\\
  \\\AND
  Filip Graliński \\
  Applica.ai \\ Warsaw, Poland \\
  \small{filip.gralinski@applica.ai}\\
  \\\And
  Tomasz Górecki \\
  Adam Mickiewicz University \\ Poznań, Poland \\
  \small{tomasz.gorecki@amu.edu.pl}\\
  \\}

%% file: abstract.tex


The paper presents a novel method of finding a fragment in a long temporal sequence similar to the set of shorter sequences. We are the first to propose an algorithm for such a search that does not rely on computing the average sequence from query examples. Instead, we use query examples as is, utilizing all of them simultaneously. The introduced method based on the Dynamic Time Warping (DTW) technique is suited explicitly for few-shot query-by-example retrieval tasks. We evaluate it on two different few-shot problems from the field of Natural Language Processing. The results show it either outperforms baselines and previous approaches or achieves comparable results when a low number of examples is available.

%% file: main.tex
\section{Introduction\label{sec:introduction}}

\begin{figure*}
    \centering
    \includegraphics[width=0.85\textwidth]{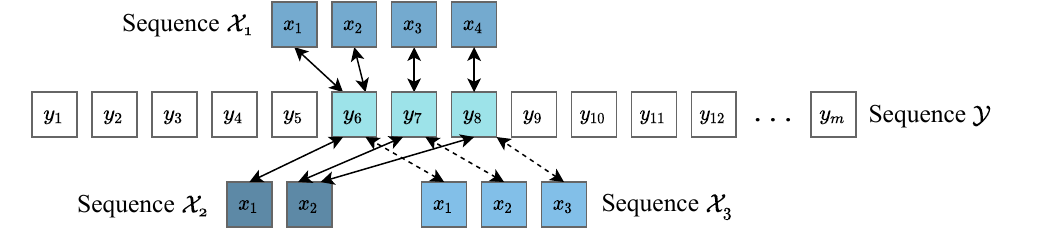}
    \caption{The problem considered is to align multiple sequences (here $\mathcal{X}_1$, $\mathcal{X}_2$, $\mathcal{X}_3$) optimally within the target sequence $\mathcal{Y}$, assuming all have to be matched to the same sub-sequence of $\mathcal{Y}$. Optimal alignment is one that minimizes the cost over all possible alignments. An example from Natural Language Processing is to locate a named entity within the sentence, given a few examples of other named entities.}
    \label{fig:idea}
\end{figure*}

This work bridges Information Retrieval, Natural Language Processing, Dynamic Programming, and Machine Learning, introducing a novel approach to identifying text spans with semantic matching. Although the method can retrieve any sequential information from an untrimmed stream, this paper demonstrates application to diverse problems involving text in natural language.

Let us start by observing that a substantial proportion of retrieval, detection, and sequence labeling tasks can be solved using sub-sequence matching. However, so far, no mainstream methods tackle the problem this way.

Consider the case of Named Entity Recognition (also referred to as entity identification, entity chunking or entity extraction, NER) -- a task of locating and classifying spans of text associated with real-world objects, such as person names, organizations, and locations, as well as with abstract temporal and numerical expressions such as dates \cite{Yadav2018ASO,Goyal2018RecentNE,Li2018ASO}.


{The problem is commonly solved with trained models for structured prediction \cite{DBLP:journals/corr/HuangXY15,DBLP:journals/corr/LampleBSKD16}. In contrast, we propose to solve it in a previously not recognized way: to use word embeddings (see Section~\ref{sec:emb}) directly}, performing semantic sub-sequence matching. In other words, determine a sentence span similar to named entities provided in the train set, with no training required beforehand. In some cases, for instance, when few-shot scenarios are considered (where only a few examples are available), this approach may be beneficial (problem was investigated in Section~\ref{sec:ner}).

Other examples can be found in the field of Information Retrieval (IR). When text documents are considered, the typical IR scenario is a provision of ranked search results for a given text query entered by a user. Search results can be either full documents or spans of texts, and each of the mentioned scenarios poses different challenges \cite{Mitra2018AnIT}.

Many modern approaches to Information Retrieval rely on a straightforward comparison of dense embeddings representing query documents and candidate documents, determining optimal results using $k$-nearest neighbor search  \cite{schmidt-etal-2019-seagle,10.1145/2983323.2983815,brokos-etal-2016-using,KIM2017122,10.1145/3196826}. When such end-to-end retrieval systems are considered, the main question becomes how to determine reliable representations of documents~\cite{gillick2018endtoend}. 

To take the approach to Information Retrieval described above, one has to already know the boundaries of units to be returned, e.g., assume sentences or paragraphs should be considered as possible results. A more challenging problem arises when we do not search for a predefined text fragment (e.g., entire document or whole sentence) but are expected to return any possible and adequate sub-sequence in a document (e.g., few sentences, several words, or even one word). This is the case for many real-world scenarios, where documents lack accessible formal structure, and one is expected to determine spans in natural language streams \cite{Vanderbeck2011AML,borchmann2020contract}. Take an example of a lawyer or researcher searching for crucial parts of legal documents to determine whether they contain fairness policies and how these policies look like~\cite{DBLP:journals/corr/abs-1809-04262}.

As shown later, it is possible to tackle the problem with a proper sub-sequence matching strategy, which can incorporate all given examples to retrieve suitable text span (Section~\ref{sec:semantic_retrieval}).



We solve the problems stated above with unconventionally used Dynamic Programming algorithms and propose their modifications. In particular, the well-known DTW Barycenter Averaging heuristic is evaluated in a~new scenario, where word embeddings are used to determine document spans. More importantly, a new sub-sequence matching method is introduced, performing a search by multiple examples simultaneously. This matching method maximizes gain from the availability of a few semantically similar text span examples. 
Because of the relation of the newly introduced method to the Dynamic Time Warping algorithm, it is referred to as the Dynamic Boundary Time Warping (DBTW\noqa{spell-DBTW}).

\begin{figure}
\centering
\includegraphics[width=0.45\textwidth]{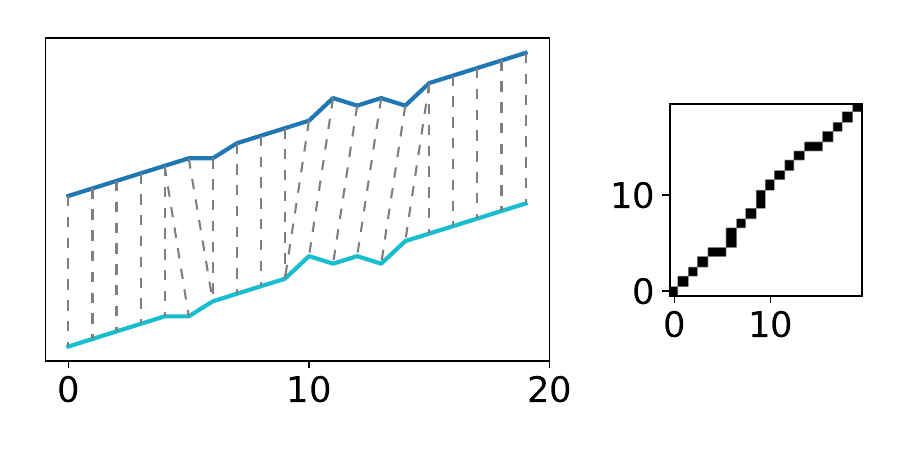}
\caption{DTW between two time series and the optimal alignment path. The dashed line connects elements aligned between up and down time series. The plot on the right depicts which time step was aligned to which, with each off-diagonal move indicating warping.\label{fig:dtw_example}}
\end{figure}



The rest of this paper is organized as follows.
Section~\ref{sec:related-work} summarizes related works in the areas of Information Retrieval, Natural Language Processing, and time-series mining. Section~\ref{sec:problem_statement} describes the problem we are dealing with.
Section~\ref{sec:dtw_algorithm_main} introduces the Dynamic Time Warping algorithm and its derivatives. 
In Section~\ref{sec:dbtw}, we present our Dynamic Boundary Time Warping algorithm together with complexity study and its adaptation to NLP problems. Section~\ref{sec:evaluation}  reports evaluation results on two different NLP tasks. Finally, Section~\ref{sec:summary} concludes the paper and outlines future research directions.

\section{Related Works\label{sec:related-work}}

Dynamic Boundary Time Warping with maximum distance limit can be considered a~binary non-parametric classifier~\cite{Boiman08indefense} over all possible document sub-sequences because it determines which of them represents the same class as positive examples. In such a~sense, its application to few-shot semantic retrieval is related to the widely studied problem of one- and few-shot learning (e.g., \citet{1597116, 1467333, Koch2015SiameseNN, Snell2017PrototypicalNF, Sung2017LearningTC}). However, these approaches are not directly comparable because, in contrast to \noqa{spell-DBTW}DBTW, knowledge obtained during training for previous categories is used.

Many time-series mining problems require subsequence similarity search as a~subroutine. While this can be performed with any distance measure, and dozens of distance measures have been proposed in the last years, there is increasing evidence that DTW is the best measure across a~wide range of domains~\cite{Ding2008}. Subsequence DTW (S-DTW) is a variant of the DTW technique~\cite{muller2007dynamic}, which is designed to find multiple similar subsequences between two templates. One of the most cited methods is SPRING~\cite{4221753}, where a~query time series is searched in a~larger streaming time series. Examples of subsequence matching applications are sensor network monitoring~\cite{4221753}, spoken keyword spotting~\cite{6418819}, sensor-based gait analysis~\cite{faucris.117708404}, acoustic~\cite{8037426}, motion capture~\cite{4812476}, or human action recognition in video~\cite{5995470}.
Additionally, to speed up computations, some hardware implementations of S-DTW-based algorithms were proposed, using GPUs and \noqa{spell-FPGAs}FPGAs~\cite{Keogh2013, 6832031, 5694075}. {Further optimizations could be achieved, e.g., by learning a kernel approximating DTW as proposed by \citet{kernel-dtw} or replacing DTW with PrunedDTW \cite{PrunedDTW}, an exact algorithm for speeding up DTW matrix calculation.}

There have been a~few attempts to utilize Dynamic Time Warping in Natural Language Processing. \citet{matuschek2008measuring} explored the earlier idea of  \citet{Ratanamahatana2004EverythingYK} to treat texts as bit streams for the purposes of measuring text similarity. \noqa{grammar-SENT_START_NUM}\citet{4338356} utilized DTW with WordNet-based word similarity to decide the semantic similarity of sentences. \citet{zhu-etal-2017-semantic} used DTW with word embeddings distances to determine the similarity between paragraphs of text to decide the similarity between whole documents. Although sub-sequence DTW was successfully applied to query-by-example tasks of spoken term detection (e.g.,~\citet{Hazen2009QuerybyexampleST, Parada2009QuerybyexampleST}), to the best of our knowledge, we are the first to apply it to plain-text query-by-example tasks. Moreover, we are unaware of any existing adaptations of sub-sequence DTW for querying by multiple examples simultaneously.



\vspace{2mm}
\begin{tcolorbox}[%
    enhanced, 
    breakable,
    frame hidden,
    ]{}
    \input{nomenclature}
\end{tcolorbox}

\section{Problem Statement\label{sec:problem_statement}}

The general problem considered is to align multiple sequences of possibly different lengths from the set $\mathbb{S}$ optimally {within} some target sequence $\mathcal{Y}$, assuming all have to be matched to the same sub-sequence of $\mathcal{Y}$ (see Figure~\ref{fig:idea}).

{The total cost of alignment between sequences from $\mathbb{S}$ and sub-sequence of the $\mathcal{Y}$ sequence is the sum of distances between all pairs of matched elements. Distance between two elements is some domain-specific measure, such as the absolute difference between scalars associated with these elements. 
Optimal alignment is one that finds such sub-sequence of $\mathcal{Y}$ that the cost of aligning all $\mathbb{S}$ within this sub-sequence is minimized over all possible sub-sequences of $\mathcal{Y}$. Sections~\ref{sec:dtw_algorithm} and \ref{sec:subsentence} provide a formal definition of the mentioned objective under additional requirements of monotonicity and continuity.}





\begin{figure*}
\centering
\includegraphics[width=0.8\textwidth,trim={6cm 0 0 0},clip]{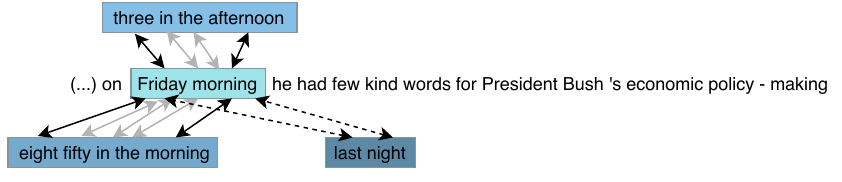}
\caption{{The DBTW matching using the semantic distance between word embeddings applied to the Named Entity Recognition problem. Here, the three examples of time expressions were matched to the \textit{Friday morning} sub-sequence.}\label{fig:real}}
\end{figure*}

An example real-word problem from Natural Language Processing is Named Entity Recognition, which may be considered under this paradigm, when one has to locate a named entity within the sentence, given a few examples of other named entities (Figure~\ref{fig:real}). Another case is semantic retrieval of legal clauses from unstructured documents, given examples of clauses covering the same topic of interest from other documents.

Note that the problem mentioned above is a generalization of every problem previously considered as a sub-sequence matching to the cases when multiple examples are available instead of a single one. Problems outside the NLP to be considered under this framework include spoken term detection or temporal activity detection in continuous, untrimmed video streams, which resembles the mentioned approach to semantic retrieval if one realizes it is in principle possible to perform sub-sequence matching on video frames.

\section{Dynamic Time Warping\label{sec:dtw_algorithm_main}}

Let us start with an introduction of a widely used Dynamic Time Warping algorithm since evaluated methods either directly use one of its variants or propose its generalization to multiple alignment scenarios. {DTW is a classical and well-established distance measure well suited to the task of comparing time series \citep{Berndt94} and was proposed by \citet{vintsyuk1968speech}.}

In general, DTW is based on the calculation of an optimal match between two given sequences, assuming one sequence is a time-warped version of another, that is, the target sequence is either stretched (one-to-many alignment), condensed (many-to-one alignment), or not warped (one-to-one alignment) concerning the source sequence (Figure~\ref{fig:dtw_example}). The optimal match is the one with the lowest cost computed as the sum of (predominantly Euclidean) distances for each matched pair of points.

\subsection{Algorithm\label{sec:dtw_algorithm}}

Classic DTW algorithm compares sequences assuming the first elements, and the last elements in both sequences are to be matched. In the case of natural language, this means that given two sentences (or documents), in every case, the first words of these will be linked with each other, as well as the last words. Although this variant is of no use in problems we consider in the present paper (see Section~\ref{sec:introduction}), there is a need to introduce it before going further.


\begin{figure}[t]
\centering
\noqa{latex-8}\includegraphics[width=0.35\textwidth]{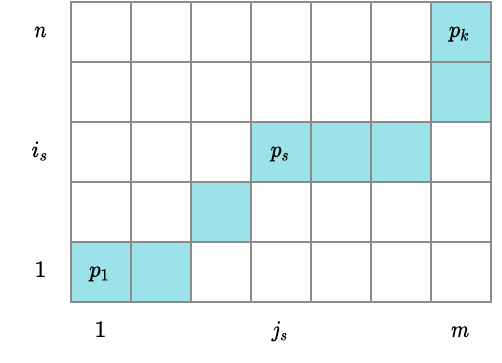}
\caption{The problem of determining the optimal match between sequences considered on $n \times m$ unit grid.\label{fig:grid1}} 
\end{figure}

The process of determining the optimal match between two time-dependent sequences $\mathcal{X} \coloneqq (x_1, \dotsi, x_n)$ and $\mathcal{Y} \coloneqq (y_1, \dotsi, y_m)$ (where $x_1, \dotsi, x_n, y_1, \dotsi, y_m$ are domain-specific objects, e.g.,~word embeddings) can be conducted on the $n \times m$ unit grid (Figure~\ref{fig:grid1}). The path through the grid $p = (p_1, \dotsi, p_s, \dotsi, p_k)$ where $p_s = (i_s, j_s)$ is referred to as the warping path, whereas the \textit{total cost} of the warping path $p$ between $\mathcal{X}$ and $\mathcal{Y}$ is given by the sum of the local cost measures for the underlying grid nodes:
\begin{align*}
\operatorname{C}_p(\mathcal{X}, \mathcal{Y}) \coloneqq \sum_{s=1}^{k} c(x_{i_s}, y_{j_s}).
\end{align*}
where $c$ is a local cost measure as defined by~\citet{muller2007dynamic}.\footnote{In Section~\ref{sec:localcost} we propose a local cost measure specifically tailored for problems in the NLP field.}

It can be further normalized with division by $n + m$, leading to the \textit{time-normalized cost}.

Let $\mathbb{P}$ denote an exponentially explosive set of all possible warping paths through the grid. The Dynamic Time Warping algorithm determines the best alignment path (\textit{optimal warping path})
\begin{align*}
 p^* = \argmin_{p \in \mathbb{P}} (\operatorname{C}_p(\mathcal{X}, \mathcal{Y}))\noqa{grammar-ENGLISH_WORD_REPEAT_RULE,latex-25}
\end{align*}
in $\mathcal{O}(nm)$ time, assuming:
\begin{itemize}
    \item the alignment path {has} to start at the bottom left of the grid ($i_1  = 1$ and $j_1 = 1$), that is the first points in both sequences are matched,
    \item monotonicity ($i_{s-1} \leq i_s$ and $j_{s-1} \leq j_s$), that is moves to the left (back in time) on the grid are not allowed,
    \item continuity ($i_s - i_{s-1} \leq 1$ and $j_s - j_{s-1} \leq 1$) that is no node on a~path can be skipped,
    \item the alignment path ends at the top right of the grid ($i_k = n$ and $j_k = m$), that is the last points in both sequences are matched,
    \item optional conditions regarding the warping window or slope constraint that can be applied in order to improve performance~\cite{Sakoe:1990:DPA:108235.108244}.
\end{itemize}


Let $D$ denote the $n \times m$ matrix referred to as the \textit{accumulated cost matrix}. The problem stated can be solved with the following initial conditions: \begin{align} \label{DTW-init}
    \begin{aligned}
      D_{i, 1} &\coloneqq \sum_{a=1}^i c(x_a, y_1),& \text{ for }& i \in \{1, \dotsi, n\}, \\
      D_{1, j} &\coloneqq \sum_{a=1}^j c(x_1, y_a),& \text{ for }& j \in \{1, \dotsi, m\}.
    \end{aligned}
\end{align}
and the following dynamic programming equation, calculated recursively in ascending order: \begin{align*}
    D_{i, j} \coloneqq c(x_i, y_j) + \min
    \left\{
        \begin{array}{ll}
          D_{i, j - 1}, \\
          D_{i - 1, j - 1}, \\
          D_{i - 1, j}.
        \end{array}
    \right.
\end{align*}

The value of $D_{n, m}$ (accumulated cost after reaching the top-right of the grid) is the total cost of the best alignment path:

\[ \operatorname{DTW}(\mathcal{X}, \mathcal{Y}) \coloneqq \operatorname{C}_{p*}(\mathcal{X}, \mathcal{Y}). \]

\subsection{Sub-sequence DTW\label{sec:subsentence}}

Mining scenarios considered in the introduction (such as Named Entity Recognition or Information Retrieval from untrimmed text streams) require slightly different behavior, offered by DTW operating on sub-sequences. It was initially introduced for problems such as the detection of spoken terms in audio recording.

In the case of sub-sequence DTW, the constraints on admissible paths are relaxed. Boundary conditions $j_1 = 1$ and $j_k = m$ are withdrawn, so the remaining $i_1 = 1$ and $i_k = n$ guarantee that the shorter sequence  $\mathcal{X}$ will be matched entirely within $\mathcal{Y}$, but not necessarily starting from the beginning of $\mathcal{Y}$ (and not obligatorily ending at the end of it). This behavior is achieved by a~modification of the initial conditions described by Equation~\eqref{DTW-init}. 
Before recursively calculating the remaining values of $D$ the first row and first column, are being set to~\cite{muller2007dynamic}: \begin{align}\label{sDTW-init}
    \begin{aligned}
        D_{i, 1}& \coloneqq \sum^i_{a=1} c(x_a, y_1),& \text{ for }& i \in \{1, \dotsi, n\},\\
        D_{1, j}& \coloneqq c(x_1, y_j),& \text{ for }& j \in \{1, \dotsi, m\}.
    \end{aligned}
\end{align}

Minimal value from the $m$th row of $D$ is the total cost of the best alignment path $\operatorname{sDTW}(\mathcal{X}, \mathcal{Y})$, whereas its index points to the $i_k$.\noqa{grammar-THE_SENT_END}



\subsection{Multi-sequence DTW}

What if one has to determine a single sub-sequence warping path for a set of short sequences? This is the case we want to consider in the present paper because this applies to few-shot semantic retrieval tasks and Named Entity Recognition. For example, it is expected to align multiple sub-sequences (named entities from train set) optimally within the target sequence (sentence or document to detect new named entities in).

\subsubsection{Exact Solution}\label{susbsub:exact}

Unfortunately, it is impossible to provide an exact solution due to practical reasons resulting from computational complexity.

\begin{figure}[t]
\centering
\includegraphics[trim={2.1cm 0 0 0},clip,width=0.32\textwidth]{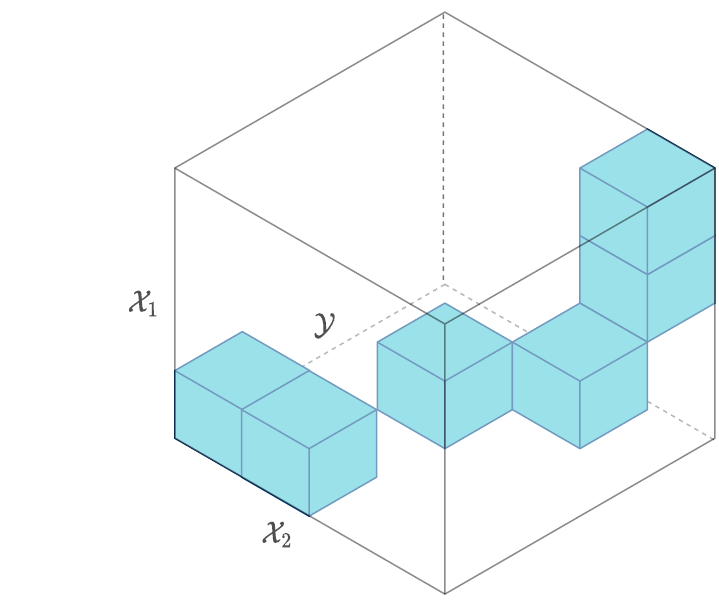}
\caption{The problem of determining the optimal match between sequences $\mathcal{X}_1, \mathcal{X}_2, \mathcal{Y}$ considered on the rectangular cuboid. Computing the optimal match would have $\mathcal{O}(n_1 n_2 m)$ time complexity.\label{fig:cuboid}}
\end{figure}

As shown by~\citet{wang1994complexity}, multiple sequence alignment with the \textit{sum of all pairs} score\footnote{When SP-score is considered, optimal alignment is one that minimizes the value over all possible alignments~\cite{bonizzoni2001complexity}.} is an~NP-complete problem. In particular, the problem of aligning $h$~sequences can be solved by applying DTW on the $h$-dimensional cuboid (see Figure~\ref{fig:cuboid}).
Assuming sequences are of the lengths $n_1, \dotsi, n_h$, the algorithm would take $\Theta(\prod_{l=1}^h n_l)$ operations and would require an exponential space, meaning that calculating it for larger $h$ is not possible in most cases~\cite{Petitjean2011AGA}.
\subsubsection{Barycenter Averaging\label{sec:barycenter}}

A reference heuristic for aligning multiple sub-sequences within the target sequence {relies on the construction of an average, consensus sequence}, representative for a given set of sentences. {The term \textit{consensus sequence} refers to a sequence which represents the most commonly encountered pattern in the set of sequences~\cite{pierce2017genetics}.}
{To approximate the optimal solution to the problem with multiple sequences, one can compute sub-sequence DTW between such consensus sequence and target sequence.}

\citet{Petitjean2011AGA} proposed the DTW Barycenter Averaging (DBA), the method for constructing consensus sequence inspired by computational biology. According to the authors, it \textit{builds an average sequence around significant states of the data}, which is \textit{truly representative of the underlying phenomenon}.

The algorithm assumes the iterative computation of an averaged sequence {(See lines 2-7 from Algorithm~\ref{alg:dba})}. Let $\mathcal{Z} = (z_1, \dotsi, z_q)$ denote the consensus sequence at the current iteration. 
First, the initial $\mathcal{Z}$ is set (e.g., as a~randomly selected element of $\mathbb{S}$). Then, during each iteration:
\begin{itemize}
    \item for each $\mathcal{X} \in \mathbb{S}$, $\operatorname{DTW}(\mathcal{X}, \mathcal{Z})$ is calculated and underlying associations\footnote{We mean DTW associations like in the Figure~\ref{fig:idea}. For example $y_6$ from Figure~\ref{fig:idea} is associated with 4 sequence's members $x_1, x_2$ from $X_1$, $x_1$ from $X_2$ and $x_1$ from $X_3$. Analogously $z_1$ from $\mathcal{Z}$ could also be associated with sequence's members from each $\mathcal{X} \in \mathbb{S}$.} resulting from the optimal warping path are stored,
    \item $\mathcal{Z}$ is updated as an average of the associated sequence's members, {e.g., word embeddings.}
\end{itemize}

During this process, the initial averaging is being refined since the new $\mathcal{Z}$ is closer to the sequences it averages concerning the total cost. {The process finishes when a new consensus sequence $\mathcal{Z}_{new}$ is almost equal to the previous consensus sequence $\mathcal{Z}_{old}$ or when the maximum number of iterations\footnote{For simplicity we omitted constraint on a number of maximum iterations criterion in Algorithm~\ref{alg:dba}.} is reached. For a thorough, detailed description of DBA, please refer to Algoritm~5 from \citet{Petitjean2011AGA}.}

{Strictly speaking,} to handle the set of sequences $\mathbb{S} = \{\mathcal{X}_1, \dotsi, \mathcal{X}_h\}$ to be aligned within $\mathcal{Y}$, one can first determine the~consensus sequence {$\mathcal{Z}^*$} from
$\mathbb{S}$ {using DBA,} and then utilize a~standard sub-sequence DTW algorithm for two sequences {(See Algorithm \ref{alg:dba})}. This approach resembles the nearest centroid classifier~\cite{10.2307/3058706} since one is determining class prototype and rely on distances between it and candidate sequences.

\algdef{SE}[DOWHILE]{Do}{doWhile}{\algorithmicdo}[1]{\algorithmicwhile\ #1}

\begin{algorithm}[t]
    \caption{DTW Barycenter Averaging based solution for aligning set of sequences $\mathbb{S}$ within target sequence $\mathcal{Y}$.\label{alg:dba}}
$\operatorname{DBA}$ is the Algorithm 5 from \citet{Petitjean2011AGA}.
\begin{algorithmic}[1]
\Procedure{MatchUsingDBA}{$\mathbb{S}, \mathcal{Y}$}
    \State $\mathcal{Z}_{new} \gets \text{random element from set } \mathbb{S}$ 
    \Do
    \State $\mathcal{Z}_{old} \gets \mathcal{Z}_{new}$
    \State $\mathcal{Z}_{new} \gets \operatorname{DBA}(\mathcal{Z}_{old}, \mathbb{S})$ 
    \doWhile{$\mathcal{Z}_{old} \not\approx \mathcal{Z}_{new}$}
    \State $\mathcal{Z}^* \gets \mathcal{Z}_{new}$
    \State \textbf{return} $\operatorname{sDTW}(\mathcal{Z}^*, \mathcal{Y})$
\EndProcedure
\end{algorithmic}
\end{algorithm}




\section{Novel Solution: Dynamic Boundary Time Warping\label{sec:dbtw}}

Contrary to the DBA, we propose a method that does not average sub-sequences before determining the best match. Simultaneously, there is a low computational cost involved, even though a form of multi-alignment is being performed.

Note that, for Information Retrieval, we are often interested only in approximating the $p^*_1$ and $p^*_k$ (more strictly the $j^*_1$ and $j^*_k$ components),\footnote{For instance, when retrieving text spans, we do not care about the alignment with the search query, but only the content (defined by $j_1$ and $j_k$).} that is the beginning and the end of the optimal warping path concerning the set of short sequences $\mathbb{S}$ and long sequence $\mathcal{Y}$. In other words, we want to find $j_1$ and $j_k$ that would minimize the sum of warping paths costs between each sequence $\mathcal{X} \in \mathbb{S}$ and the long sequence $\mathcal{Y}$:
\begin{equation*}
    j^*_1, j^*_k = \argmin_{j_1, j_k}\Big(\sum_{\mathcal{X} \in \mathbb{S}} \operatorname{C}_p(\mathcal{X}, \mathcal{Y})\Big).
\end{equation*}
Note that the final warping paths between considered sequences have the same $j^*_1$, $j^*_k$.
Calculating such optimal solution is more straightforward than presented in Section~\ref{susbsub:exact}, but still too time-consuming for long sequence $\mathcal{Y}$, because one would have to consider all possible $j_1$ and $j_k$ pairs (see Section~\ref{subsub:complexity}). 
The situation changes when we allow either $j_1$ or $j_k$ to be different among examined warping paths, for instance, as it will be shown later (see Algorithm~\ref{alg:multidtw1}), we can easily find $$\displaystyle j^*_k = \argmin_{j_k}\Big(\sum_{\mathcal{X} \in \mathbb{S}} \operatorname{C}_p(\mathcal{X}, \mathcal{Y})\Big).$$ Our algorithm exploits this fact, and searches for the $j_k$ first ($j_1$ being unconstrained), and then for $j_1$ given previously determined optimal $j_k$.
We will use the name Dynamic Boundary Time Warping to highlight this difference when referring to the proposed solution.



Let us introduce the generalized $\operatorname{DTW}$ (or $\operatorname{gDTW}$) first. We will use this term when referring to the $\operatorname{DTW}$ that is parameterized by the pre-initialized accumulated cost matrix $D$.
For example, for $D$ initialized from Equation~\eqref{DTW-init}:
\begin{equation*}
\operatorname{gDTW}(\mathcal{X}, \mathcal{Y}, D_{\eqref{DTW-init}}) = DTW(\mathcal{X}, \mathcal{Y})
\end{equation*}
and
for $D$ initialized from Equation~\eqref{sDTW-init}:
\begin{equation*}
\operatorname{gDTW}(\mathcal{X}, \mathcal{Y}, D_{\eqref{sDTW-init}}) = sDTW(\mathcal{X}, \mathcal{Y}).
\end{equation*}

DBTW\noqa{spell-DBTW} degenerates to \noqa{spell-sDTW}sDTW in the case of $|\mathbb{S}| = 1$, that is when only one example is available. The complete computation when multiple examples are given is detailed in Algorithm~\ref{alg:multidtw1} and Algorithm~\ref{alg:multidtw2}. We propose to handle the problem as follows: \begin{itemize}
    \item Initialize the accumulated cost matrix $D$ from Equation~\eqref{sDTW-init} for each of the $\mathbb{S}$ elements independently.
    \item Calculate $\operatorname{sDTW}$ for each of the $\mathbb{S}$ elements independently, time-normalize underlying accumulated cost matrices, and sum their $m$-th rows. The result can be used to determine $p^*_k = (i^*_k, j^*_k)$ analogously to the conventional sub-sequence DTW.
    \item Reverse $\mathcal{Y}$, as well as all sequences in $\mathbb{S}$, and initialize $D'$ for each reversed sequence from $\mathbb{S}$: \begin{equation}
      \label{ourDTW-init}
      \begin{gathered}[b]
        D'_{i, 1} \coloneqq \sum^i_{a=1} c(x'_a, y'_1)  \quad \text{for } i \in \{1, \dots, n\}, \\
        D'_{1, j} \coloneqq \infty \quad  \text{ for } j \in \{1, \dots,m\} \setminus {j'^*_1}, \\
        D'_{1, j'^*_1} \coloneqq c(x_1, y_{j'^*_1}), 
      \end{gathered}
    \end{equation}
    where $j'^*_1 = m - j^*_k + 1$.
    \item Calculate $\operatorname{gDTW}$ (using $D'$)
    on reversed sequences with the constraint that it should start with $p'^*_1 = (1, m - j^*_k + 1$), that is $p^*_k$ after reversal. In this way $p'^*_k$ is determined, which gives $p^*_1 = (1, m - j'^*_k + 1)$, that is $p'^*_k$ after reversal. 
\end{itemize}
Note that DBTW\noqa{spell-DBTW} first finds an optimal, common $j^*_k$ for all sequences in $\mathbb{S}$ (starting indexes could be different). Then, all sequences are reversed, and $j'^*_k$ is determined by forcing the algorithm to start from $j'^*_1$. This way, such $j^*_1$ and $j^*_k$ are found that approximate an optimal solution.

\begin{algorithm}[t]
    \caption{Approximation of optimal $j_k$ for the multiple sub-sequences DTW problem.\label{alg:multidtw1}}
    \begin{algorithmic}[1]
    \small
    \Procedure{MultiWarpingEnd}{$\mathbb{S}, \mathcal{Y}, equation$}
    \State $\Vec{sum}\gets (0, \dotsi, 0)$
    \For{$l\gets 1, |\mathbb{S}|$}
        \State $D^l\gets D^l$ from $equation$
       \State $\operatorname{gDTW}(\mathcal{X}_l, \mathcal{Y}, D^l)$
       \State $\Vec{sum}\gets \Vec{sum}+D_{n, *}^l$
    \EndFor
    \State $j_k \gets \argmin_i(\Vec{sum}_i)$
    \State \textbf{return} $j_k$
    \EndProcedure
    \end{algorithmic}
\end{algorithm}

\begin{algorithm}[t]
    \begin{algorithmic}[1]
    \small
    \Procedure{Rev}{$\mathcal{X}$} \Comment{Sequence $(x_1, \dotsi, x_n)$}
       \State \textbf{return} $(x_n, x_{n-1}, \dotsi, x_1)$
    \EndProcedure
    \\
    \Procedure{MatchUsingDBTW}{$\mathbb{S}, \mathcal{Y}$}
    \State $j_k \gets \Call{MultiWarpingEnd}{\mathbb{S}, \mathcal{Y}, \text{Equation~\eqref{sDTW-init}}}$
    \State $\mathcal{Y}' \gets \Call{Rev}{\mathcal{Y}}$
    \State $\mathbb{S}' \gets \{\Call{Rev}{\mathcal{X}} : \mathcal{X} \in \mathbb{S} \} $
    \State $j_k' \gets \Call{MultiWarpingEnd}{\mathbb{S}', \mathcal{Y}', \text{Equation~\eqref{ourDTW-init}}}$
    \State $j_1 \gets m - j_k' + 1$
    \State \textbf{return} $j_1, j_k$
    \EndProcedure
    \end{algorithmic}
    \caption{Approximation of optimal $j_1$ and $j_k$ for the multiple sub-sequences DTW problem.\label{alg:multidtw2}}
\end{algorithm}

\subsection{Complexity Study}\label{subsub:complexity}
Let us assume that the set of short sequences $\mathbb{S}$ consists of $h$ sequences of length $n$, and long sequence $\mathcal{Y}$ is of length $m$.

DBA based solution from Algorithm~\ref{alg:dba} consists of two parts: (1) calculation of consensus sequence using DBA, and (2) calculation of sDTW between consensus sequence and $\mathcal{Y}$ sequence.

As described by \citet{Petitjean2011AGA}, the time complexity of Step~1 is equal to $\Theta(bn^2h)$, where $b$ refers to the number of iterations needed for DBA to converge.
Since the complexity of Step~2 is $\Theta(nm)$, the complexity of all steps is equal to $\Theta(bn^2h + nm)$.

The most costly operation for DBTW is the \Call{MultiWarpingEnd}{} procedure, which for each sequence in $\mathbb{S}$ computes gDTW with $\mathcal{Y}$ sequence, and it is called twice. Therefore DBTW time complexity is equal to $\Theta(2nmh)=\Theta(nmh)$.

Depending on the problem setup, the time complexity of DBTW can be either smaller or higher than the complexity of the DBA solution.

Note that the optimal solution requires to compute gDTW between $\mathcal{Y}$ and each sequence in $\mathbb{S}$ for every possible $j_1$ and $j_k$. Since there are $\frac{m(m+1)}{2}$ such possible unique pairs of $j_1$ and $j_k$, the overall complexity is equal to $\Theta(nmh \times \frac{m(m+1)}{2})=\Theta(nm^3h)$, which is larger than the time complexity of DBTW and in most common cases larger than the DBA solution's complexity.

\subsection{Local Cost for Natural Language Processing Problems\label{sec:localcost}}

There is a need to propose a suitable local cost function to apply any DTW-based dynamic programming algorithms to problems from the field of Natural Language Processing. We introduce a novel approach, relying on the distance between contextualized word embeddings.

\subsubsection{Contextualized Word Embeddings\label{sec:emb}}

Roughly speaking, the reasoning behind word embeddings is to follow the distributional hypothesis, according to which \textit{difference of meaning correlates with the difference of distribution} \cite{doi:10.1080/00437956.1954.11659520}. This means words sharing context tend to share similar meanings, and one is able to obtain semantic representations of words by optimizing some auxiliary objective in a sizeable unlabeled text corpus.

A famous example is the Continuous Bag of Words (CBOW) model, where an average of vectors representing surrounding words is used as an input to log-linear classifier predicting the target (middle) word \cite{Mikolov2013EfficientEO}. This simple yet effective algorithm and the skip-gram model trained with the opposite objective have taken the world of word embeddings by storm \cite{Young2018RecentTI}.

Representations provided using CBOW and similar models, however, are static. This means that when the pre-trained word embeddings are used in a downstream task, the representation of a given word is context-invariant: \textit{wound} used as a past tense of \textit{wind} share representation with \textit{wound} denoting \textit{to injure}.

Later approaches of  \citet{DBLP:journals/corr/abs-1802-05365}, and \citet{akbik2018coling} assume the use of deep language models' internal states. These, contrary to static word embeddings, are expected to capture context-dependent word semantics. Resulting contextualized word embeddings are a~function of the entire input sentence, such as for a~sequence of $z$ input tokens, an associated sequence of $z$ vectors is returned.

Early contextualized word embeddings were sourced from language models using Recurrent Neural Networks, and they are currently being replaced by language models based on the architecture of Transformers~\cite{DBLP:journals/corr/VaswaniSPUJGKP17} such as BERT~\cite{DBLP:journals/corr/abs-1810-04805}, GPT-2~\cite{gpt2}, or RoBERTa~\cite{liu2019roberta}. In~the case of embeddings sourced from Transformer-based language models, the representation is obtained by attending to different tokens of the input sentence~\cite{Ethayarajh2019HowCA}.

To the best of our knowledge, only \citet{zhu-etal-2017-semantic} used Dynamic Time Warping with word embeddings, and none of the previous attempts were based on contextualized word embeddings.

\subsubsection{Distance Measure}

Many distance measures may be applied as local cost functions. In some domains, simple distance measures such as Euclidean distance are sufficient enough \cite{ShiehKeogh08}, whereas in other, it may be beneficial to use learned distance metric \cite{7953240}. 

In the case of Natural Language Processing, we propose to rely on the cosine distance between contextualized word embeddings as the local cost, which is defined as: \begin{align*}
    c(\pmb x, \pmb y) = \frac{1 - {
        \frac {\pmb x \cdot \pmb y}{||\pmb x||\, ||\pmb y||}
    }}{2}.
\end{align*}
where, $||\pmb x||$ is $\ell_2$-norm, and $\pmb x \cdot \pmb y$ is the dot product of the two vectors.

It is the most common metric used in NLP tasks when dissimilarity between two word vectors is considered \cite{faruqui-etal-2016-problems}.







\subsubsection{Optional Weighting}

Methods of determining document similarity tend to benefit from the inclusion of frequency or distribution information, such as in Inverse Document Frequency~\cite{Metzler2008GeneralizedID} or Smooth Inverse Frequency (SIF) weighting ~\cite{arora2017asimple}. We propose to further extend the algorithm with the additional weight factor $w$ applied to the DTW equation: \begin{align*}
    D_{i, j} \coloneqq w_i \cdot c(x_i, y_j) + \min
    \left\{
        \begin{array}{ll}
          D_{i, j - 1},\\
          D_{i - 1, j - 1},\\
          D_{i - 1, j}.
        \end{array}
    \right.
\end{align*}
The $w_i$ is defined as the SIF of the underlying token $t_i$: 
\begin{align*}
    w_i^{SIF} = \frac{a}{a + f_i},
\end{align*}
where $f_i$ stands for relative frequency of the token $t_i$ and $a$ is the weight parameter, recommended\noqa{grammar-ADMIT_ENJOY_VB} to be between $10^{-3}$ and $10^{-4}$~\cite{arora2017asimple}.

The intuition behind the introduction of such weighting is to capture the importance of the token when calculating an accumulated cost, in such a~way that less informative (more probable) words contribute less to the final score.

\subsection{Implementation Details}

The performance of local cost calculations is the primary factor when one is bound by time or resource restrictions in the case of DTW and similar algorithms \cite{Myers1979PerformanceTI}. Since a cosine distance between word embeddings is used in our scenario, there is a need to calculate at least $n \times m$ distances (for the one-shot scenario) between vectors of 768 or more components, where $n$ denote the number of words in positive example and $m$ stands for the length of the document.

We were able to compute them efficiently with GPU and CUDA parallel computing platform. In our PyTorch-based implementation  \cite{NEURIPS2019_9015} for given input matrices representing embeddings of sequences to compare, a matrix of cosine distances is returned. It is further cast to NumPy array \cite{oliphant2006guide} used in the Dynamic Programming part, which is implemented using Numba (JIT compiler translating Python and NumPy code into fast machine code, see \citet{10.1145/2833157.2833162}).

\section{Evaluation\label{sec:evaluation}}

The introduced Dynamic Boundary Time Warping algorithm has broad applications in few-shot retrieval tasks from a~variety of domains. We restricted ourselves to already established problems within the field of Natural Language Processing. For these, simple albeit specialized proof-of-concept solutions were provided.

In each setting, an addition to DBTW has been proposed to facilitate handling the specific problem and demonstrate the algorithm's extensibility.

\subsection{Few-shot Semantic Retrieval\label{sec:semantic_retrieval}}

The recently proposed contract discovery task \citep{borchmann2020contract} aims to provide spans of requested target documents semantically similar to examples of spans from a few other documents. The mentioned dataset is intended to test the mechanisms that detect legal texts' regulations, given a few examples of other clauses regulating the same issue (query-by-multiple-examples scenario). Sample spans often vary in length, and the contained text is written using different vocabulary or syntax. Moreover, the text to search in lacks a formal structure, that is, no segmentation into distinct sections, articles, paragraphs, or points is given in advance.

For example, given two examples of text, where the parties agree on which jurisdiction the contract will be subject to:

\begin{displayquote}
\small
This Agreement shall be governed by and construed under the laws of the State of California without reference to its rules of conflicts of laws.
\end{displayquote}

\begin{displayquote}
\small
This Agreement is governed by the internal laws of the State of Florida and may be modified or waived only in writing signed by the Party against which such modification or waiver is sought to be enforced.
\end{displayquote}

\noindent match the following text span in another document: 

\begin{displayquote}
\small
Each party hereto consents to exclusive personal jurisdiction in the State of Delaware and voluntarily submits to the jurisdiction of the courts of the State of Delaware in any action or proceeding concerning this Agreement.
\end{displayquote}

Because each word is represented by word embedding that reflects its meaning, and we can compute the distance between any pair of embeddings (Section~\ref{sec:localcost}), it is in principle possible to state that California is semantically quite similar to Delaware.

As a result, it is possible to attempt matching clauses such as the two shown above into the third one -- word by word, embedding by embedding. Due to this fact, the problem of contract discovery is suited for the DBTW algorithm -- it can be perceived as an alignment of multiple sequences (examples of desirable text spans from other legal documents) optimally within the target sequence (document in which one wants to determine a text span regulating the same issue).

Contract Discovery is evaluated with Soft F$_1$ metric calculated on character-level spans, as implemented in \textit{GEval} tool~\cite{gralinski-etal-2019-geval}. Roughly speaking, this is the conventional $F_{1}$ measure, with precision and recall definitions altered to reflect the partial success of returning entities. As a result, identifying half of the correct span does not result in a 0 score.

\paragraph{Experiment.} DBA and Adaptive CBOW solutions were evaluated in addition to DBTW. All utilized the same finetuned GPT\nobreakdash-1 model, as described by \citet{borchmann2020contract}. We decided to utilize GPT\nobreakdash-1 instead of GPT\nobreakdash-2 because the authors achieved comparable results for both of them. At the same time, the latter has more parameters, larger embeddings, and more fine-grained tokenization, while all of these have a significant performance impact.

{The GPT-1 Language Model we used was originally introduced by \citet{Radford2018ImprovingLU} who proposed to rely on the decoder of multi-layer Transformer \cite{vaswani2017attention}. The authors released a 12-layer model with 768-dimensional states and 12 attention heads. It uses a BPE vocabulary \cite{sennrich-etal-2016-neural} consisting of 40,000 sub-word units. \citet{borchmann2020contract} fine-tuned the model for 40 epochs on a corpus of legal documents, using a standard, next-word prediction objective. The authors used the initial learning rate of $5e-5$, linear learning rate decay, and Adam optimizer with decoupled weight decay \cite{loshchilov2019decoupled}.\footnote{Both model and the corpus are publicly available at \url{http://github.com/applicaai/contract-discovery}}
We used internal states from the last layer of the model as word embeddings, leading to the dimensionality of 768.
}

Because of the annotation assumptions made in this shared task, it is often beneficial to return the whole sentence, even though one can find the exact location of the desired clause (within the sentence). {Consider an example of the following sentence:
\begin{displayquote}
\small
This Agreement shall be governed by and construed and enforced in accordance with the laws of the State of Georgia...

...as to all matters regardless of the laws that might otherwise govern under principles of conflicts of laws applicable thereto.
\end{displayquote}
Here, DBTW selects only the first part, and it would be desirable to highlight it for an end user in the real-world application. Nevertheless, it was preferred to keep the complete sentence as an expected clause during the preparation of \citet{borchmann2020contract} dataset. The annotator selected an incomplete sentence only when the remaining, non-important part was of a greater length than the crucial one, which contains the desired information.
} That is the reason why we were returning results rounded in order to match the entire sentence that ``clause core'' was found in. 

\paragraph{Baseline.}
In Algorithm~\ref{alg:acbow} {we introduce} the Adaptive Continuous Bag of Words (ACBOW), a~simple and fast algorithm, 
that represents a~straightforward, natural approach to tackling the problem. Roughly speaking, the idea is to move with a~constantly changing window over tokens from~$\mathcal{Y}$ and determine the best sub-sequence (Algorithm~\ref{alg:search_acbow}). Embeddings for each text fragment are averaged and the resulting vectors compared with cosine similarity.
In the case of multiple sequences, an average of individual similarities to the considered window is used (procedure $\operatorname{SIM}$ in Algorithm~\ref{alg:search_acbow}).

Note that the ACBOW for which the results were reported in Table~\ref{tab:legal} differs from the ACBOW Algorithm~\ref{alg:acbow}. The former was extended with a~possibility to look into the future and check if adding more tokens would improve an overall score, even when some of them temporarily lower the similarity.

\begin{algorithm}[t]
    \begin{algorithmic}[1]
    \small
    \Procedure{SIM}{$\mathbb{S}, u$}
    \State $\mathbb{E} \gets \{\Call{mean}{\mathcal{X}} : \mathcal{X} \in \mathbb{S}\}$
    \State $e_u \gets \Call{mean}{u}$
    \State $scores \gets \{c(e_u, e) : e \in \mathbb{E}\}$
    \State \textbf{return} $\Call{mean}{scores}$
    \EndProcedure
    \\
    \Procedure{FindOne}{$\mathbb{S}, \mathcal{Y}, j$}
    \State $u^* \gets (y_j)$
    \State $u \gets ()$
    \While{$j + 1 \leq m \textbf{ and } u \ne u^*$}
        \State $u \gets u^*$
        \State $u' \gets (u^*_1, \ldots, u^*_r, y_{j+1})$
        \If{$\Call{sim}{\mathbb{S}, u^*} < \Call{sim}{\mathbb{S}, u'}$}
            \State $u^* \gets u'$
            \State $j \gets j + 1$
        \EndIf
        \State $u' \gets (u^*_2, \ldots, u^*_r)$
        \If{$ u' \ne () \textbf{ and } \Call{sim}{\mathbb{S}, u^*} < \Call{sim}{\mathbb{S}, u'}$}
            \State $u^* \gets u'$
        \EndIf
    \EndWhile
    \State \textbf{return} $u^*, \Call{sim}{\mathbb{S}, u}, j+1$
    \EndProcedure
    \end{algorithmic}
    \caption{Finding one similar sub-sequence $u = (u_1, \ldots, u_r)$ from $\mathcal{Y}$ to $\mathbb{S}$ sequences given starting index $j$.\label{alg:search_acbow}}
\end{algorithm}

\begin{algorithm}[t]
    \begin{algorithmic}[1]
    \small
    \Procedure{MatchUsingACBOW}{$\mathbb{S}, \mathcal{Y}$}
    \State $j \gets 1$
    \While{$j \leq m$}
    \State $u, score, j \gets \Call{FindOne\noqa{spell-FindOne}}{\mathbb{S}, \mathcal{Y}, j}$
        \If{$score > score^*$}
            \State $u^*, score^* \gets u, score$
        \EndIf
    \EndWhile
    \State \textbf{return} $u^*, score^*$
    \EndProcedure
    \end{algorithmic}
    \caption{Finding most similar subsequence $u = (u_1, \ldots, u_r)$ from $\mathcal{Y}$ given $\mathbb{S}$ sequences using ACBOW algorithm.\label{alg:acbow}}
\end{algorithm}

\paragraph{Results.} Table~\ref{tab:legal} summarizes the Soft F$_1$ scores achieved. Contrary to what one might suspect, the Adaptive CBOW baseline was unable to provide satisfactory results. Scores of the sub-sequence DTW with a~DBA-determined consensus sequence were substantially higher. The usage of cosine distance instead of Euclidean seems beneficial in the case of DBA used with word embeddings. DBTW performs the best, and its effectiveness can be attributed to both inverse frequency weighting and the proposed way of handling multiple sequences. The new method proposed in this paper slightly outperforms the method presented by \citet{borchmann2020contract} even when fICA projection\footnote{\citet{borchmann2020contract} used decomposition of contextualized word embeddings based on Independent Component Analysis \cite{Hyvrinen2000IndependentCA} and observed it helps to distinguish semantically differing texts. See Table~\ref{tab:legal} for comparison.} of embeddings was not applied. It is worth mentioning that SIF weighting does not lead to an improvement in the aforementioned paper. Results were even better when both SIF and fICA projection was used.

{There are several distinguishing features the improvement over \citet{borchmann2020contract} can be attributed to. First of all, there is a reduction of noise that occurs in DBTW. Recall the example of the governing law clause presented at the beginning of Section. The first part of the sentence contains information required to correctly classify the clause, whereas the rest is a potential noise source. The DBTW considers all the possible sub-sentences and is not restricted to the sentence boundaries, as is the method proposed by \citet{borchmann2020contract}. Secondly, DBTW is not order-invariant, and thus it can easily capture key phrases and word n-grams. Thirdly, DBTW operates on word-level, whereas other methods rely on averaged representation of multiple, possibly a few hundred words. The latter results in yet additional noise and information loss.} 

Moreover, note that \citet{borchmann2020contract} chose the most similar spans from the sentence n-grams. Although their approach leads to comparable results to those obtained with DBTW, it could be applied to a limited number of problems when the number of considered n-grams is low. In contrast, DBTW is not subject to such constraints and can effectively search for a very long sequence. For example, when word-level (instead of sentence-level) sequences are considered, they often become much longer, and the n-gram based methods would be too expensive computationally.

\begin{table}
    \centering
    \small
    \caption{Results of solutions based on the same finetuned GPT\nobreakdash-1 model as described by \citet{borchmann2020contract}, obtained on test set.\label{tab:legal}} 
    \begin{tabularx}{0.3\textwidth}{lr} \toprule
    Method & Soft F$_{1}$ \\ \midrule
    \citet{borchmann2020contract} & \\
    \quad $-$fICA & $.47$ \\
    \vspace{3pt} \quad $+$fICA & $.49$ \\
    \vspace{3pt} ACBOW & $.35$ \\
    DBA & \\
    \quad Euclidean & .43 \\
    \vspace{3pt} \quad Cosine & $.44$ \\
    DBTW & \\
    \quad $-$SIF & .47 \\
    \quad $+$SIF ($a=10^{-3}$) & $.50$ \\
    \quad $+$SIF \, $+$fICA & $\pmb{.51}$ \\
    \bottomrule
    \end{tabularx}
\end{table}

{Most of the mentioned advantages also apply to the DBA. However, one may hypothesize that information loss occurring during the consensus sequence calculation is substantial in long passages from the Contract Discovery dataset. Similarly, ACBOW shares some desired properties of DBTW (e.g., consideration of arbitrary sub-sequence on word-level) but, contrary to the DBTW, is order-invariant and relies on noisy averaged representations of multiple word embeddings.}




\subsection{Few-shot Named Entity Recognition\label{sec:ner}}

Named Entity Recognition is the task of tagging entities in text with their corresponding type. These differ depending on the dataset. In the case of the richly-annotated Ontonotes corpus~\cite{pradhan-etal-2013-towards}, tags such as people and organization names, locations, languages, events, monetary values, and more are used.

There were several attempts to the NER problem in a few-shot scenario \cite{Fritzler:2019:FCN:3297280.3297378,hofer2018fewshot}. Since the mentioned setting is in line with our problem statement (Section~\ref{sec:problem_statement}), we approached it to provide another proof-of-concept from the field of NLP. As outlined in Section~\ref{sec:introduction}, we solve the problem of Named Entity Recognition with a new approach of semantic sub-sequence matching.

Named Entity Recognition task differs substantially from Semantic Retrieval discussed in the previous section. To tackle the problem effectively, one has to notice there is a~significant variance in lengths of entities to be retrieved—they can range from one word to over a~dozen words within the same class. This fact could motivate non-trivial modifications of DBTW such as:
\begin{itemize}
    \item Normalization of accumulated costs for sequences from $\mathbb{S}$ in order to compensate the impact of longer sequences on the overall score (otherwise the longer individual warping path is, the higher would be its impact when choosing the approximately optimal path for the set of sequences).
    \item Preference for either contraction or expansion when determining the warping path for a single sequence, e.g., depending on its length in relation to average named entity length.
\end{itemize}

There are multiple normalization methods to consider in the former, whereas the latter may require the introduction of warping path bands to restrict the upper length of matched sub-sequence. We decided to take a more straightforward, which solves both problems at the same time:
\begin{itemize}
    \item Given the set of sequences $\mathbb{S}$, take the length of the longest as a~target size. 
    \item Resample shorter sequences to reach the target size using interpolation with the spline of order 1, as implemented in tslearn \textit{TimeSeriesResampler}~\cite{tslearn}.
\end{itemize}
After this step, no further normalization nor weights adjustments may be required to provide satisfactory results.

{Because the number of results to be returned for a given sentence varies from zero to few, one cannot simply return the most similar sub-sequence in the case of Named Entity Recognition. We tackle the problem by introducing a threshold and return all non-overlapping paths from the given sentence, with an accumulated cost below the assumed distance level. Given a set of training examples $\mathbb{S}$, we calculate DBTW($\mathbb{S} \setminus \{\mathcal{X}\}$, $\mathcal{X}$) for each $\mathcal{X} \in \mathbb{S}$.
The threshold is calculated as the maximal cost of optimal warping path from such inner-train matches. The threshold for DBA is determined analogously.
}

\paragraph{Experiment.} We roughly followed the procedure for evaluation of a few-shot NER\noqa{grammar-CHILDISH_LANGUAGE} proposed by \citet{Fritzler:2019:FCN:3297280.3297378}. Authors trained models on subsamples of Ontonotes development set~\cite{pradhan-etal-2013-towards} for each class separately.\footnote{The original train set was used as a source of \textit{out-of-domain} data in part of scenarios, but this does not apply to methods based on DBTW. Similarly, as a~baseline, we relied on an approach, which utilizes only \textit{in-domain} training data. See \citet{Fritzler:2019:FCN:3297280.3297378} for details regarding this distinction.}
For each case, $h=20$ sentences containing a particular named entity were selected. Besides, sentences without considered entity had all the classes replaced with \texttt{O}, and part of them were added to the train set, to preserve the original distribution of the currently evaluated class. Note that $h$ is not necessarily equal to the number of annotations available since it is common for one Ontonotes sentence to contain more than one named entity of the same type.

In our case, solutions were evaluated for $h \in [1, 10]$, since we are aiming mainly at good performance for a lower number of examples available. Moreover, ten experiments with different random seeds were conducted for each class, instead of four performed by \citet{Fritzler:2019:FCN:3297280.3297378}.

\begin{figure*}
\includegraphics[width=\textwidth]{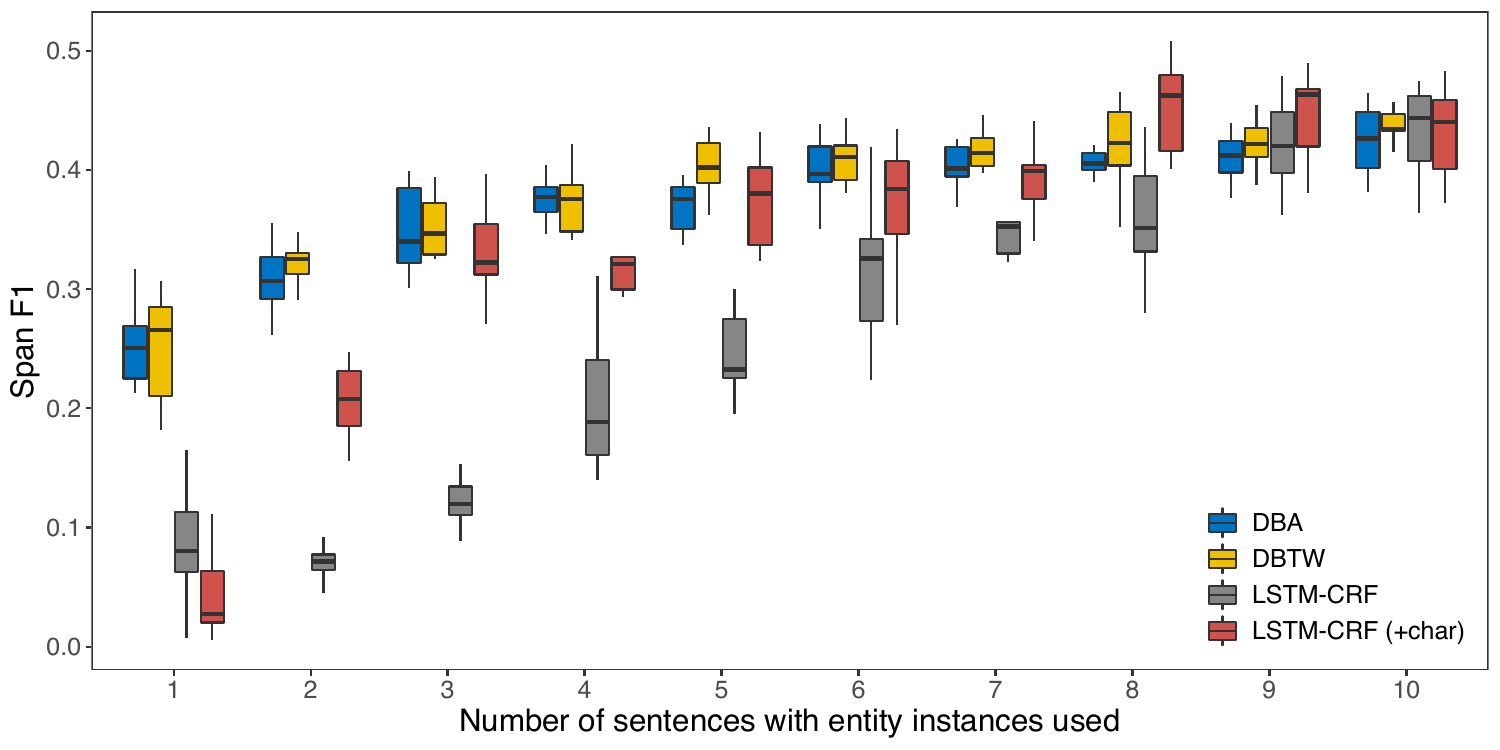}
\caption{Performance in Named Entity Recognition as a~function of the number of sentences with positive examples available. {Note that LSTM-CRF ($+$char) model is not directly comparable because, contrary to the LSTM-CRF, DBA, and DBTW, it uses character-level embeddings in addition to ELMo and GloVe.}
\label{fig:ner}}
\end{figure*}

\paragraph{Baseline.} LSTM-CRF used as a~reference is a~BiLSTM-CRF model trained on ELMo and GloVe embeddings. It follows the specification of \citet{Fritzler:2019:FCN:3297280.3297378}, but with the difference that trained character embeddings were not used to simplify the comparison with DBTW. Note that otherwise, one had to propose a~procedure of training character embeddings compatible with DBTW, which is beyond the scope of this paper. {Nevertheless, we report results of LSTM-CRF with trained character-level embeddings for the sake of completeness.}

{The remaining LSTM-CRF baseline, DBA, and DBTW approaches rely on the same embeddings, resulting from the concatenation of the 1024-dimensional ELMo model released by \citet{Peters:2018} with the original 50-dimensional GloVe embeddings \cite{Pennington14glove:global}.
Although \citet{Fritzler:2019:FCN:3297280.3297378} trained their baselines for 20 epochs, we found our models undertrained in this setting and decided to enlarge the value to 30 epochs.}

\begin{table*}[t]
    \centering
    \caption{{$p$-values for permutation t-test comparing DBA and DBTW.}}
    \label{tab:p_values1}
    {
    \begin{tabular}{lcccccccccc}
    \toprule
     $n$ & 1 & 2 & 3 & 4 & 5 & 6 & 7 & 8 & 9 & 10\\ 
     \midrule
     $p$-value & 0.9339 & 0.3895 & 0.8779 & 0.8803 & 0.0038 & 0.4499 & 0.309 & 0.2049 & 0.2161 & 0.1727\\
     \bottomrule
    \end{tabular}}
\end{table*}

\begin{table*}[t]
    \centering
    \caption{{$p$-values for permutation t-test comparing DBTW and LSTM-CRF (+char).}}
    \label{tab:p_values2}
    {
    \begin{tabular}{lcccccccccc}
    \toprule
     $n$ & 1 & 2 & 3 & 4 & 5 & 6 & 7 & 8 & 9 & 10\\ 
     \midrule
     $p$-value & 0.0001 & 0.0001 & 0.1262 & 0.0025 & 0.0606 & 0.0482 & 0.1114 & 0.0693 & 0.0819 & 0.7024\\
     \bottomrule
    \end{tabular}}
\end{table*}

\paragraph{Results.} Comparison of DBTW, DBA, and LSTM-CRF with the same input embeddings is presented on Figure~\ref{fig:ner}. \textit{Span F1} score refers to a commonly used $F_{\beta=1}$ variant where exact matches of the corresponding entities are considered~\cite{10.3115/1119176.1119195}.

Both DBA and DBTW outperform the LSTM-CRF baseline in a few-shot setting. Noteworthy, DTW-based methods receive near-identical scores in the experiment. {In order to statistically compare methods, we decided to use the permutation t-test. The implemented test corresponds to the proposal of \citet{Chung13}. While a permutation test requires that we see all possible permutations of the data (which can become quite large), we can easily conduct ``approximate permutation tests'' by simply conducting a very large number of samples (we used 10,000 permutations instead of 3,628,800 possible permutations). That process should, in expectation, approximate the permutation distribution. Obtained $p$-values we can find in Table~\ref{tab:p_values1} and Table~\ref{tab:p_values2}}.

{From Table~\ref{tab:p_values1} we can see that it is possible to reject ($\alpha = 5\%$) the null hypothesis (about equality of methods DBA and DBTW) only for $n=5$ (the same we can read from Figure~\ref{fig:ner}). In such situations, it seems reasonable to assume that methods do not differ significantly. 

Comparable results of DBTW and DBA can be potentially attributed to two factors. Firstly, named entities in ontonotes are usually short: 58\% of the test set entities consist of a single word and 21\% -- of two words. When one-word sub-sequences are to be considered, the methods are roughly equivalent. We expect DBTW to perform better in the case of long sequences because it is where noise related to the calculation of the DBA consensus sequence emerges. Secondly, we found the problem of determining the number of sub-sequences to return, which occurs in both DBA and DBTW, to play an important role. If the sentence contains a named entity of a particular type, the highest-scored sub-sequence can be classified as such with high confidence. E.g., we can maximize recall by withdrawing the threshold and returning the top result. Nevertheless, precision suffers without the threshold, and the simple heuristics we experimented with are unable to provide an optimal cut-off.}

{LSTM-CRF with character-level embeddings seems to converge faster than the LSTM-CRF baseline. It appears that it achieves scores comparable to DBTW for five and more sentences in the train set (Table~\ref{tab:p_values2}). However, due to the reasons outlined at the beginning, the methods cannot be directly compared.}

\section{Summary and Future Work\label{sec:summary}}

In this paper, an algorithm inspired by Dynamic Time Warping was proposed, as well as a new application of existing DBA Barycenter Averaging heuristics. 
It was shown how to adapt it to current problems in the field of Natural Language Processing as a~result of cosine distance applied to contextualized word embeddings.
Unlike its predecessors, Dynamic Boundary Time Warping can find an approximate solution for the problem of querying by multiple examples. What is crucial, the proposed approach is in some applications substantially better than calculating a~consensus sequence and utilizing it to perform sub-sequence DTW search, presumably because there is no unnecessary information loss involved. Due to the inclusion of inverse frequency weighting specific to NLP problems, its effectiveness was further improved. Thus it was able to outperform methods previously proposed for \textit{Few-shot Contract Discovery} with the same Language Model applied.


Applications of the proposed algorithm are not limited to the cases where proof-of-concept solutions were provided, and it can be applied to other few-shot retrieval tasks. 
Problems outside the NLP to be considered under this framework include temporal activity detection in continuous, untrimmed video streams~\cite{cMontes, Xu2019TwoStreamRC}, which resembles mentioned approach to Semantic Retrieval if one realizes it is in principle possible to perform sub-sequence matching on video frame embeddings. Such can be encoded with a pretrained image classification network (i.e., ResNeXt \cite{Xie2016AggregatedRT}) and processed analogously. Moreover, the DBTW applies to every problem previously considered as a sub-sequence matching when multiple examples are available instead of a single one.

\section*{Acknowledgements}

The Smart Growth Operational Programme supported this research under project no. POIR.01.01.01-00-0605/19 (\textit{Disruptive adoption of Neural Language Modelling for automation of text-intensive work}).


%% file: nomenclature.tex
\renewcommand{\nomname}{List of Symbols}
\mbox{}
\nomenclature{$\mathcal{Y}$}{Time-dependent target sequence; \\ $\mathcal{Y} \coloneqq (y_1, \dotsi, y_m)$}
\nomenclature{$\mathcal{X}$}{Time-dependent sequence to align within target sequence $\mathcal{Y}$; \\ $\mathcal{X} \coloneqq (x_1, \dotsi, x_n)$}
\nomenclature{$n$}{Length of sequence $\mathcal{X}$}
\nomenclature{$n_l$}{Length of sequence $\mathcal{X}_l$}
\nomenclature{$m$}{Length of sequence $\mathcal{Y}$}
\nomenclature{$k$}{Length of warping path $p$}
\nomenclature{$i$}{Index of $i$th element of $\mathcal{X}$}
\nomenclature{$j$}{Index of $j$th element of $\mathcal{Y}$}
\nomenclature[a$j1$]{$j^{*}_{1}$}{Index of the beginning of optimal sub-sequence alignment in $\mathcal{Y}$}
\nomenclature[a$j2$]{$j'^{*}_{1}$}{Index of the beginning of optimal sub-sequence alignment in $\mathcal{Y'}$; $j'^{*}_{1} = m - j^{*}_{k} + 1$}
\nomenclature[a$j1$]{$j^{*}_{k}$}{Index of the end of optimal sub-sequence alignment in $\mathcal{Y}$}
\nomenclature[a$j2$]{$j'^{*}_{k}$}{Index of the end of optimal sub-sequence alignment in $\mathcal{Y'}$; $j'^{*}_{k} = m - j^{*}_{1} + 1$}
\nomenclature{$w$}{Additional weight factor applied to the DTW
equation}
\nomenclature{$\mathcal{Z}$}{Consensus sequence at the current iteration; $\mathcal{Z} \coloneqq (z_1, ..., z_q)$}
\nomenclature{$\mathcal{Z}^*$}{Final consensus sequence}
\nomenclature{$u$}{Sub-sequence from $\mathcal{Y}$ similar to sequences from set $\mathbb{S}$; $u \coloneqq (u_1,\dotsi,u_r)$}
\nomenclature{$u^*$}{Sub-sequence from $\mathcal{Y}$ most similar to sequences from set $\mathbb{S}$}
\nomenclature{$e$}{Element of set $\mathbb{E}$}
\nomenclature{$e_u$}{Embedding representing sequence $u$}
\nomenclature{$\mathbb{E}$}{Set of embeddings, each embedding represent different sequence from set $\mathbb{S}$}
\nomenclature{$r$}{Length of the $u$ sub-sequence}
\nomenclature{$f_i$}{Relative frequency of the token $t_i$}
\nomenclature{$t_i$}{$i$th token corresponding to $i$th element of $\mathcal{X}$}
\nomenclature{$a$}{Hyperparameter of the smooth inverse frequency (SIF) method}
\nomenclature{$b$}{Number of iterations needed for DTW Barycenter Averaging (DBA) to converge}
\nomenclature{$s$}{Index of $s$th element of warping path $p$}
\nomenclature{$l$}{Index of $l$th element of set $\mathbb{S}$}
\nomenclature{$p$}{Warping path; $p \coloneqq (p_1, \dotsi, p_s, \dotsi, p_k)$}
\nomenclature[a$p0$]{$p^*$}{Optimal warping path; \\ $p^* \coloneqq \argmin_{p \in \mathbb{P}} (\operatorname{C}_p(\mathcal{X}, \mathcal{Y}))$}
\nomenclature[a$p1$]{$p^*_1$}{First element of optimal warping path in $D$; $p^*_1 = (1, j^*_1)$}
\nomenclature[a$p2$]{$p'^*_1$}{First element of optimal warping path in $D'$; $p'^*_1 = (1, j'^*_1)$}
\nomenclature[a$p1$]{$p^*_k$}{Last element of optimal warping path in $D$; $p^*_k = (n, j^*_k)$}
\nomenclature[a$p2$]{$p'^*_k$}{Last element of optimal warping path in $D'$; $p'^*_k = (n, j'^*_k)$}
\nomenclature{$x_i, y_j$}{Domain-specific objects e.g., word embeddings}
\nomenclature[a$c2$]{$\operatorname{C}_p(\mathcal{X}, \mathcal{Y})$}{Cost of the warping path $p$ between $\mathcal{X}$ and $\mathcal{Y}$; $\operatorname{C}_p(\mathcal{X}, \mathcal{Y}) \coloneqq \sum_{s=1}^{k} c(x_{i_s}, y_{j_s})$}
\nomenclature[a$c1$]{$c(x_i, y_j)$}{Local cost measure for domain-specific objects $x_i$ and $y_j$ e.g., cosine distance between word embeddings}
\nomenclature{$\mathbb{P}$}{Exponentially explosive set of all possible warping paths through the grid}
\nomenclature{$D$}{Accumulated cost matrix of size $n \times m$ calculated from $\mathcal{X}$, $\mathcal{Y}$}
\nomenclature{$D'$}{Accumulated cost matrix of size $n \times m$ calculated from $\mathcal{X}'$, $\mathcal{Y}'$}
\nomenclature{$D_{i, j}^l$}{Item from $i$th row and $j$th column of matrix $D$ calculated from $\mathcal{X}_l$,  $\mathcal{Y}$}
\nomenclature{$h$}{Size of set $\mathbb{S}$}
\nomenclature{$q$}{Length of sequence $\mathcal{Z}$}
\nomenclature{$\mathbb{S}$}{Set of time-depended sequences $\mathbb{S}$; \\ $\mathbb{S} \coloneqq \{\mathcal{X}_1, \dotsi, \mathcal{X}_h\}$}
\nomenclature{$\mathcal{X}'$}{Reversed sequence of $X$; \\ $\mathcal{X'} \coloneqq (x_n, \dotsi, x_1) = (x'_1, \dotsi, x'_n)$}
\nomenclature{$\mathcal{Y}'$}{Reversed sequence of $Y$; \\ $\mathcal{Y'} \coloneqq (y_m, \dotsi, y_1) = (y'_1, \dotsi, y'_m)$}

\printnomenclature

%% file: dtw-for-few-shot.bbl
\begin{thebibliography}{84}
\expandafter\ifx\csname natexlab\endcsname\relax\def\natexlab#1{#1}\fi

\bibitem[{Akbik et~al.(2018)Akbik, Blythe, and Vollgraf}]{akbik2018coling}
Alan Akbik, Duncan Blythe, and Roland Vollgraf. 2018.
\newblock Contextual string embeddings for sequence labeling.
\newblock In \emph{{COLING} 2018, 27th International Conference on
  Computational Linguistics}, pages 1638--1649.

\bibitem[{Arora et~al.(2017)Arora, Liang, and Ma}]{arora2017asimple}
Sanjeev Arora, Yingyu Liang, and Tengyu Ma. 2017.
\newblock \href {https://openreview.net/forum?id=SyK00v5xx} {A simple but
  tough-to-beat baseline for sentence embeddings}.
\newblock In \emph{5th International Conference on Learning Representations,
  {ICLR} 2017, Toulon, France, April 24-26, 2017, Conference Track
  Proceedings}. OpenReview.net.

\bibitem[{{Bart} and {Ullman}(2005)}]{1467333}
Evgeniy {Bart} and Shimon {Ullman}. 2005.
\newblock \href {https://doi.org/10.1109/CVPR.2005.117} {Cross-generalization:
  learning novel classes from a single example by feature replacement}.
\newblock In \emph{2005 IEEE Computer Society Conference on Computer Vision and
  Pattern Recognition (CVPR'05)}, volume~1, pages 672--679 vol. 1.

\bibitem[{Barth et~al.(2015)Barth, Oberndorfer, Pasluosta, Schülein, Gaßner,
  Reinfelder, Kugler, Schuldhaus, Winkler, Klucken, and
  Eskofier}]{faucris.117708404}
Jens Barth, Cäcilia Oberndorfer, Cristian~Federico Pasluosta, Samuel
  Schülein, Heiko Gaßner, Samuel Reinfelder, Patrick Kugler, Dominik
  Schuldhaus, Jürgen Winkler, Jochen Klucken, and Björn Eskofier. 2015.
\newblock \href {https://doi.org/10.3390/s150306419} {{Stride} {Segmentation}
  {During} {Free} {Walk} {Movements} {Using} {Multi}-dimensional {Subsequence}
  {Dynamic} {Time} {Warping} on {Inertial} {Sensor} {Data}}.
\newblock \emph{Sensors}, 15:6419--6440.
\newblock UnivIS-Import:2015-04-14:Pub.2015.tech.IMMD.IMMD5.stride.

\bibitem[{Berndt and Clifford(1994)}]{Berndt94}
Donald~J. Berndt and James Clifford. 1994.
\newblock Using dynamic time warping to find patterns in time series.
\newblock In \emph{Proceedings of the 3rd International Conference on Knowledge
  Discovery and Data Mining}, AAAIWS'94, pages 359--370. AAAI Press.

\bibitem[{Boiman et~al.(2008)Boiman, Shechtman, and Irani}]{Boiman08indefense}
Oren Boiman, Eli Shechtman, and Michal Irani. 2008.
\newblock In defense of nearest-neighbor based image classification.
\newblock In \emph{In IEEE Conference on Computer Vision and Pattern
  Recognition (CVPR}, pages 1--8.

\bibitem[{Bonizzoni and Della~Vedova(2001)}]{bonizzoni2001complexity}
Paola Bonizzoni and Gianluca Della~Vedova. 2001.
\newblock The complexity of multiple sequence alignment with sp-score that is a
  metric.
\newblock \emph{Theoretical Computer Science}, 259(1-2):63--79.

\bibitem[{Borchmann et~al.(2020)Borchmann, Wiśniewski, Gretkowski, Kosmala,
  Jurkiewicz, Szałkiewicz, Pałka, Kaczmarek, Kaliska, and
  Graliński}]{borchmann2020contract}
Łukasz Borchmann, Dawid Wiśniewski, Andrzej Gretkowski, Izabela Kosmala,
  Dawid Jurkiewicz, Łukasz Szałkiewicz, Gabriela Pałka, Karol Kaczmarek,
  Agnieszka Kaliska, and Filip Graliński. 2020.
\newblock \href {http://arxiv.org/abs/1911.03911} {Contract discovery: Dataset
  and a few-shot semantic retrieval challenge with competitive baselines}.

\bibitem[{Boytsov et~al.(2016)Boytsov, Novak, Malkov, and
  Nyberg}]{10.1145/2983323.2983815}
Leonid Boytsov, David Novak, Yury Malkov, and Eric Nyberg. 2016.
\newblock \href {https://doi.org/10.1145/2983323.2983815} {Off the beaten path:
  Let’s replace term-based retrieval with k-nn search}.
\newblock In \emph{Proceedings of the 25th ACM International on Conference on
  Information and Knowledge Management}, CIKM ’16, page 1099–1108, New
  York, NY, USA. Association for Computing Machinery.

\bibitem[{Brokos et~al.(2016)Brokos, Malakasiotis, and
  Androutsopoulos}]{brokos-etal-2016-using}
Georgios-Ioannis Brokos, Prodromos Malakasiotis, and Ion Androutsopoulos. 2016.
\newblock \href {https://doi.org/10.18653/v1/W16-2915} {Using centroids of word
  embeddings and word mover{'}s distance for biomedical document retrieval in
  question answering}.
\newblock In \emph{Proceedings of the 15th Workshop on Biomedical Natural
  Language Processing}, pages 114--118, Berlin, Germany. Association for
  Computational Linguistics.

\bibitem[{Candelieri et~al.(2019)Candelieri, Fedorov, and Messina}]{kernel-dtw}
Antonio Candelieri, Stanislav Fedorov, and Vincenzina Messina. 2019.
\newblock \href {https://doi.org/10.3390/s19235192} {Efficient kernel-based
  subsequence search for enabling health monitoring services in iot-based home
  setting}.
\newblock \emph{Sensors}, 19:5192.

\bibitem[{{Chen} et~al.(2009){Chen}, {Chen}, {Chen}, and Ooi}]{4812476}
Yueguo {Chen}, Gang {Chen}, Ke~{Chen}, and Beng~Chin Ooi. 2009.
\newblock \href {https://doi.org/10.1109/ICDE.2009.20} {Efficient processing of
  warping time series join of motion capture data}.
\newblock In \emph{2009 IEEE 25th International Conference on Data
  Engineering}, pages 1048--1059.

\bibitem[{Chung and Romano(2013)}]{Chung13}
EunYi Chung and Joseph~P. Romano. 2013.
\newblock Exact and asymptotically robust permutation tests.
\newblock \emph{The Annals of Statistics}, 41(2):484--507.

\bibitem[{Devlin et~al.(2018)Devlin, Chang, Lee, and
  Toutanova}]{DBLP:journals/corr/abs-1810-04805}
Jacob Devlin, Ming{-}Wei Chang, Kenton Lee, and Kristina Toutanova. 2018.
\newblock \href {http://arxiv.org/abs/1810.04805} {{{BERT:} Pre-training of
  Deep Bidirectional Transformers for Language Understanding}}.
\newblock \emph{CoRR}, abs/1810.04805.

\bibitem[{Ding et~al.(2008)Ding, Trajcevski, Scheuermann, Wang, and
  Keogh}]{Ding2008}
Hui Ding, Goce Trajcevski, Peter Scheuermann, Xiaoyue Wang, and Eamonn Keogh.
  2008.
\newblock \href {https://doi.org/10.14778/1454159.1454226} {Querying and mining
  of time series data: Experimental comparison of representations and distance
  measures}.
\newblock \emph{Proc. VLDB Endow.}, 1(2):1542–1552.

\bibitem[{Ethayarajh(2019)}]{Ethayarajh2019HowCA}
Kawin Ethayarajh. 2019.
\newblock {How Contextual are Contextualized Word Representations? Comparing
  the Geometry of BERT, ELMo, and GPT-2 Embeddings}.
\newblock \emph{ArXiv}, abs/1909.00512v1.

\bibitem[{Faruqui et~al.(2016)Faruqui, Tsvetkov, Rastogi, and
  Dyer}]{faruqui-etal-2016-problems}
Manaal Faruqui, Yulia Tsvetkov, Pushpendre Rastogi, and Chris Dyer. 2016.
\newblock \href {https://doi.org/10.18653/v1/W16-2506} {Problems with
  evaluation of word embeddings using word similarity tasks}.
\newblock In \emph{Proceedings of the 1st Workshop on Evaluating Vector-Space
  Representations for {NLP}}, pages 30--35, Berlin, Germany. Association for
  Computational Linguistics.

\bibitem[{Fritzler et~al.(2019)Fritzler, Logacheva, and
  Kretov}]{Fritzler:2019:FCN:3297280.3297378}
Alexander Fritzler, Varvara Logacheva, and Maksim Kretov. 2019.
\newblock \href {https://doi.org/10.1145/3297280.3297378} {Few-shot
  classification in named entity recognition task}.
\newblock In \emph{Proceedings of the 34th ACM/SIGAPP Symposium on Applied
  Computing}, SAC '19, pages 993--1000, New York, NY, USA. ACM.

\bibitem[{Gillick et~al.(2018)Gillick, Presta, and Tomar}]{gillick2018endtoend}
Daniel Gillick, Alessandro Presta, and Gaurav~Singh Tomar. 2018.
\newblock \href {http://arxiv.org/abs/1811.08008} {{End-to-End Retrieval in
  Continuous Space}}.

\bibitem[{Goyal et~al.(2018)Goyal, Gupta, and Kumar}]{Goyal2018RecentNE}
Archana Goyal, Vishal Gupta, and Manish Kumar. 2018.
\newblock Recent named entity recognition and classification techniques: A
  systematic review.
\newblock \emph{Comput. Sci. Rev.}, 29:21--43.

\bibitem[{Grali{\'n}ski et~al.(2019)Grali{\'n}ski, Wr{\'o}blewska,
  Stanis{\l}awek, Grabowski, and G{\'o}recki}]{gralinski-etal-2019-geval}
Filip Grali{\'n}ski, Anna Wr{\'o}blewska, Tomasz Stanis{\l}awek, Kamil
  Grabowski, and Tomasz G{\'o}recki. 2019.
\newblock \href {https://www.aclweb.org/anthology/W19-4826} {{GE}val: Tool for
  debugging {NLP} datasets and models}.
\newblock In \emph{Proceedings of the 2019 ACL Workshop BlackboxNLP: Analyzing
  and Interpreting Neural Networks for NLP}, pages 254--262, Florence, Italy.
  Association for Computational Linguistics.

\bibitem[{Guo et~al.(2012)Guo, Huang, and Zhao}]{6418819}
Hongyu Guo, Dongmei Huang, and Xiaoqun Zhao. 2012.
\newblock \href {https://doi.org/10.1109/ICNIDC.2012.6418819} {An algorithm for
  spoken keyword spotting via subsequence dtw}.
\newblock In \emph{2012 3rd IEEE International Conference on Network
  Infrastructure and Digital Content}, pages 573--576.

\bibitem[{Gysel et~al.(2018)Gysel, de~Rijke, and Kanoulas}]{10.1145/3196826}
Christophe~Van Gysel, Maarten de~Rijke, and Evangelos Kanoulas. 2018.
\newblock \href {https://doi.org/10.1145/3196826} {Neural vector spaces for
  unsupervised information retrieval}.
\newblock \emph{ACM Trans. Inf. Syst.}, 36(4).

\bibitem[{{Gündoğdu} and {Saraçlar}(2017)}]{7953240}
Batuhan {Gündoğdu} and Murat {Saraçlar}. 2017.
\newblock \href {https://doi.org/10.1109/ICASSP.2017.7953240} {Distance metric
  learning for posteriorgram based keyword search}.
\newblock In \emph{2017 IEEE International Conference on Acoustics, Speech and
  Signal Processing (ICASSP)}, pages 5660--5664.

\bibitem[{Harris(1954)}]{doi:10.1080/00437956.1954.11659520}
Zellig~S. Harris. 1954.
\newblock \href {https://doi.org/10.1080/00437956.1954.11659520}
  {Distributional structure}.
\newblock \emph{WORD}, 10(2-3):146--162.

\bibitem[{Hazen et~al.(2009)Hazen, Shen, and White}]{Hazen2009QuerybyexampleST}
Timothy~J. Hazen, Wade Shen, and Christopher~M. White. 2009.
\newblock Query-by-example spoken term detection using phonetic posteriorgram
  templates.
\newblock \emph{2009 IEEE Workshop on Automatic Speech Recognition \&
  Understanding}, pages 421--426.

\bibitem[{Hoai et~al.(2011)Hoai, Lan, and la~Torre}]{5995470}
Minh Hoai, Zhen-Zhong Lan, and Fernando~De la~Torre. 2011.
\newblock \href {https://doi.org/10.1109/CVPR.2011.5995470} {Joint segmentation
  and classification of human actions in video}.
\newblock In \emph{CVPR 2011}, pages 3265--3272.

\bibitem[{Hofer et~al.(2018)Hofer, Kormilitzin, Goldberg, and
  Nevado-Holgado}]{hofer2018fewshot}
Maximilian Hofer, Andrey Kormilitzin, Paul Goldberg, and Alejo Nevado-Holgado.
  2018.
\newblock \href {http://arxiv.org/abs/1811.05468} {Few-shot learning for named
  entity recognition in medical text}.

\bibitem[{Huang et~al.(2013)Huang, Dai, Sun, Wang, Wang, and Yang}]{6832031}
Sitao Huang, Guohao Dai, Yuliang Sun, Zilong Wang, Yu~Wang, and Huazhong Yang.
  2013.
\newblock \href {https://doi.org/10.1109/HPCC.and.EUC.2013.149} {{DTW}-based
  subsequence similarity search on amd heterogeneous computing platform}.
\newblock In \emph{2013 IEEE 10th International Conference on High Performance
  Computing and Communications 2013 IEEE International Conference on Embedded
  and Ubiquitous Computing}, pages 1054--1063.

\bibitem[{Huang et~al.(2015)Huang, Xu, and Yu}]{DBLP:journals/corr/HuangXY15}
Zhiheng Huang, Wei Xu, and Kai Yu. 2015.
\newblock \href {http://arxiv.org/abs/1508.01991} {Bidirectional {LSTM-CRF}
  models for sequence tagging}.
\newblock \emph{CoRR}, abs/1508.01991.

\bibitem[{Hyv{\"a}rinen and Oja(2000)}]{Hyvrinen2000IndependentCA}
Aapo Hyv{\"a}rinen and Erkki Oja. 2000.
\newblock Independent component analysis: algorithms and applications.
\newblock \emph{Neural networks : the official journal of the International
  Neural Network Society}, 13 4-5:411--30.

\bibitem[{Kim et~al.(2017)Kim, Fiorini, Wilbur, and Lu}]{KIM2017122}
Sun Kim, Nicolas Fiorini, W.~John Wilbur, and Zhiyong Lu. 2017.
\newblock \href {https://doi.org/https://doi.org/10.1016/j.jbi.2017.09.014}
  {Bridging the gap: Incorporating a semantic similarity measure for
  effectively mapping pubmed queries to documents}.
\newblock \emph{Journal of Biomedical Informatics}, 75:122 -- 127.

\bibitem[{Koch et~al.(2015)Koch, Zemel, and Salakhutdinov}]{Koch2015SiameseNN}
Gregory Koch, Richard Zemel, and Ruslan Salakhutdinov. 2015.
\newblock Siamese neural networks for one-shot image recognition.
\newblock In \emph{Proceedings of the 32nd International Conference on Machine
  Learning}, Lille, France.

\bibitem[{Lam et~al.(2015)Lam, Pitrou, and Seibert}]{10.1145/2833157.2833162}
Siu~Kwan Lam, Antoine Pitrou, and Stanley Seibert. 2015.
\newblock \href {https://doi.org/10.1145/2833157.2833162} {Numba: A llvm-based
  python jit compiler}.
\newblock In \emph{Proceedings of the Second Workshop on the LLVM Compiler
  Infrastructure in HPC}, LLVM ’15, New York, NY, USA. Association for
  Computing Machinery.

\bibitem[{Lample et~al.(2016)Lample, Ballesteros, Subramanian, Kawakami, and
  Dyer}]{DBLP:journals/corr/LampleBSKD16}
Guillaume Lample, Miguel Ballesteros, Sandeep Subramanian, Kazuya Kawakami, and
  Chris Dyer. 2016.
\newblock \href {http://arxiv.org/abs/1603.01360} {Neural architectures for
  named entity recognition}.
\newblock \emph{CoRR}, abs/1603.01360.

\bibitem[{Li et~al.(2018)Li, Sun, Han, and Li}]{Li2018ASO}
Jing Li, Aixin Sun, Jianglei Han, and Chenliang Li. 2018.
\newblock A survey on deep learning for named entity recognition.
\newblock \emph{ArXiv}, abs/1812.09449.

\bibitem[{{Li Fei-Fei} et~al.(2006){Li Fei-Fei}, {Fergus}, and
  {Perona}}]{1597116}
{Li Fei-Fei}, R.~{Fergus}, and P.~{Perona}. 2006.
\newblock \href {https://doi.org/10.1109/TPAMI.2006.79} {One-shot learning of
  object categories}.
\newblock \emph{IEEE Transactions on Pattern Analysis and Machine
  Intelligence}, 28(4):594--611.

\bibitem[{{Liu} et~al.(2007){Liu}, {Zhou}, and {Zheng}}]{4338356}
X.~{Liu}, Y.~{Zhou}, and R.~{Zheng}. 2007.
\newblock \href {https://doi.org/10.1109/ICSC.2007.48} {{Sentence Similarity
  based on Dynamic Time Warping}}.
\newblock In \emph{International Conference on Semantic Computing (ICSC 2007)},
  pages 250--256.

\bibitem[{Liu et~al.(2019)Liu, Ott, Goyal, Du, Joshi, Chen, Levy, Lewis,
  Zettlemoyer, and Stoyanov}]{liu2019roberta}
Yinhan Liu, Myle Ott, Naman Goyal, Jingfei Du, Mandar Joshi, Danqi Chen, Omer
  Levy, Mike Lewis, Luke Zettlemoyer, and Veselin Stoyanov. 2019.
\newblock \href {http://arxiv.org/abs/1907.11692} {{RoBERTa: A Robustly
  Optimized BERT Pretraining Approach}}.

\bibitem[{Loshchilov and Hutter(2019)}]{loshchilov2019decoupled}
Ilya Loshchilov and Frank Hutter. 2019.
\newblock \href {http://arxiv.org/abs/1711.05101} {Decoupled weight decay
  regularization}.

\bibitem[{Matuschek et~al.(2008)Matuschek, Schl{\"u}ter, and
  Conrad}]{matuschek2008measuring}
Michael Matuschek, Tim Schl{\"u}ter, and Stefan Conrad. 2008.
\newblock Measuring text similarity with dynamic time warping.
\newblock In \emph{Proceedings of the 2008 international symposium on Database
  engineering \& applications}, volume 299, pages 263--267. ACM.

\bibitem[{Metzler(2008)}]{Metzler2008GeneralizedID}
Donald Metzler. 2008.
\newblock {Generalized Inverse Document Frequency}.
\newblock In \emph{CIKM}.

\bibitem[{Mikolov et~al.(2013)Mikolov, Chen, Corrado, and
  Dean}]{Mikolov2013EfficientEO}
Tomas Mikolov, Kai Chen, Gregory~S. Corrado, and Jeffrey Dean. 2013.
\newblock Efficient estimation of word representations in vector space.
\newblock \emph{CoRR}, abs/1301.3781.

\bibitem[{Mitra and Craswell(2018)}]{Mitra2018AnIT}
Bhaskar Mitra and Nick Craswell. 2018.
\newblock An introduction to neural information retrieval.
\newblock \emph{Found. Trends Inf. Retr.}, 13:1--126.

\bibitem[{Montes et~al.(2016)Montes, Salvador, Pascual-deLaPuente, and
  i~Nieto}]{cMontes}
Alberto Montes, Amaia Salvador, Santiago Pascual-deLaPuente, and
  Xavier~Gir{\'o} i~Nieto. 2016.
\newblock Temporal activity detection in untrimmed videos with recurrent neural
  networks.
\newblock In \emph{1st NIPS Workshop on Large Scale Computer Vision Systems
  2016}.

\bibitem[{M{\"u}ller(2007)}]{muller2007dynamic}
Meinard M{\"u}ller. 2007.
\newblock {Dynamic Time Warping}.
\newblock \emph{{Information Retrieval for Music and Motion}}, pages 69--84.

\bibitem[{{Myers} et~al.(1980){Myers}, {Rabiner}, and
  {Rosenberg}}]{Myers1979PerformanceTI}
Cory {Myers}, Lawrence {Rabiner}, and Aaron {Rosenberg}. 1980.
\newblock Performance tradeoffs in dynamic time warping algorithms for isolated
  word recognition.
\newblock \emph{IEEE Transactions on Acoustics, Speech, and Signal Processing},
  28(6):623--635.

\bibitem[{Nagpal et~al.(2018)Nagpal, Wadhwa, Gupta, Shaikh, Mehta, and
  Goyal}]{DBLP:journals/corr/abs-1809-04262}
Rashmi Nagpal, Chetna Wadhwa, Mallika Gupta, Samiulla Shaikh, Sameep Mehta, and
  Vikram Goyal. 2018.
\newblock \href {http://arxiv.org/abs/1809.04262} {Extracting fairness policies
  from legal documents}.
\newblock \emph{CoRR}, abs/1809.04262.

\bibitem[{Oliphant(2006)}]{oliphant2006guide}
Travis Oliphant. 2006.
\newblock \emph{A guide to NumPy}, volume~1.
\newblock Trelgol Publishing USA.

\bibitem[{Parada et~al.(2009)Parada, Sethy, and
  Ramabhadran}]{Parada2009QuerybyexampleST}
Carolina Parada, Abhinav Sethy, and Bhuvana Ramabhadran. 2009.
\newblock {Query-by-example Spoken Term Detection For OOV terms}.
\newblock \emph{2009 IEEE Workshop on Automatic Speech Recognition \&
  Understanding}, pages 404--409.

\bibitem[{Paszke et~al.(2019)Paszke, Gross, Massa, Lerer, Bradbury, Chanan,
  Killeen, Lin, Gimelshein, Antiga, Desmaison, Kopf, Yang, DeVito, Raison,
  Tejani, Chilamkurthy, Steiner, Fang, Bai, and Chintala}]{NEURIPS2019_9015}
Adam Paszke, Sam Gross, Francisco Massa, Adam Lerer, James Bradbury, Gregory
  Chanan, Trevor Killeen, Zeming Lin, Natalia Gimelshein, Luca Antiga, Alban
  Desmaison, Andreas Kopf, Edward Yang, Zachary DeVito, Martin Raison, Alykhan
  Tejani, Sasank Chilamkurthy, Benoit Steiner, Lu~Fang, Junjie Bai, and Soumith
  Chintala. 2019.
\newblock \href
  {http://papers.neurips.cc/paper/9015-pytorch-an-imperative-style-high-performance-deep-learning-library.pdf}
  {Pytorch: An imperative style, high-performance deep learning library}.
\newblock In H.~Wallach, H.~Larochelle, A.~Beygelzimer, F.~d\textquotesingle
  Alch\'{e}-Buc, E.~Fox, and R.~Garnett, editors, \emph{Advances in Neural
  Information Processing Systems 32}, pages 8024--8035. Curran Associates, Inc.

\bibitem[{Pennington et~al.(2014)Pennington, Socher, and
  Manning}]{Pennington14glove:global}
Jeffrey Pennington, Richard Socher, and Christopher~D. Manning. 2014.
\newblock {Glove: Global vectors for word representation}.
\newblock In \emph{In EMNLP}.

\bibitem[{Peters et~al.(2018{\natexlab{a}})Peters, Neumann, Iyyer, Gardner,
  Clark, Lee, and Zettlemoyer}]{DBLP:journals/corr/abs-1802-05365}
Matthew~E. Peters, Mark Neumann, Mohit Iyyer, Matt Gardner, Christopher Clark,
  Kenton Lee, and Luke Zettlemoyer. 2018{\natexlab{a}}.
\newblock \href {http://arxiv.org/abs/1802.05365} {Deep contextualized word
  representations}.
\newblock \emph{CoRR}, abs/1802.05365.

\bibitem[{Peters et~al.(2018{\natexlab{b}})Peters, Neumann, Iyyer, Gardner,
  Clark, Lee, and Zettlemoyer}]{Peters:2018}
Matthew~E Peters, Mark Neumann, Mohit Iyyer, Matt Gardner, Christopher Clark,
  Kenton Lee, and Luke Zettlemoyer. 2018{\natexlab{b}}.
\newblock Deep contextualized word representations.
\newblock In \emph{Proc. of NAACL}.

\bibitem[{Petitjean et~al.(2011)Petitjean, Ketterlin, and
  Gançarski}]{Petitjean2011AGA}
François Petitjean, Alain Ketterlin, and Pierre Gançarski. 2011.
\newblock A global averaging method for dynamic time warping, with applications
  to clustering.
\newblock \emph{Pattern Recognition}, 44:678--693.

\bibitem[{Pierce(2017)}]{pierce2017genetics}
Benjamin Pierce. 2017.
\newblock \emph{Genetics. A Conceptual Approach}.
\newblock W. H. Freeman.

\bibitem[{Pradhan et~al.(2013)Pradhan, Moschitti, Xue, Ng, Bj{\"o}rkelund,
  Uryupina, Zhang, and Zhong}]{pradhan-etal-2013-towards}
Sameer Pradhan, Alessandro Moschitti, Nianwen Xue, Hwee~Tou Ng, Anders
  Bj{\"o}rkelund, Olga Uryupina, Yuchen Zhang, and Zhi Zhong. 2013.
\newblock \href {https://www.aclweb.org/anthology/W13-3516} {Towards robust
  linguistic analysis using {O}nto{N}otes}.
\newblock In \emph{Proceedings of the Seventeenth Conference on Computational
  Natural Language Learning}, pages 143--152, Sofia, Bulgaria. Association for
  Computational Linguistics.

\bibitem[{Radford et~al.(2018)Radford, Narasimhan, Salimans, and
  Sutskever}]{Radford2018ImprovingLU}
Alec Radford, Karthik Narasimhan, Tim Salimans, and Ilya Sutskever. 2018.
\newblock Improving language understanding with unsupervised learning.

\bibitem[{Radford et~al.(2019)Radford, Wu, Child, Luan, Amodei, and
  Sutskever}]{gpt2}
Alec Radford, Jeffrey Wu, Rewon Child, David Luan, Dario Amodei, and Ilya
  Sutskever. 2019.
\newblock \href
  {https://d4mucfpksywv.cloudfront.net/better-language-models/language-models.pdf}
  {Language models are unsupervised multitask learners}.
\newblock Technical report, OpenAI.

\bibitem[{Rakthanmanon et~al.(2013)Rakthanmanon, Campana, Mueen, Batista,
  Westover, Zhu, Zakaria, and Keogh}]{Keogh2013}
Thanawin Rakthanmanon, Bilson Campana, Abdullah Mueen, Gustavo Batista, Brandon
  Westover, Qiang Zhu, Jesin Zakaria, and Eamonn Keogh. 2013.
\newblock \href {https://doi.org/10.1145/2500489} {Addressing big data time
  series: Mining trillions of time series subsequences under dynamic time
  warping}.
\newblock \emph{ACM Trans. Knowl. Discov. Data}, 7(3).

\bibitem[{Ratanamahatana and Keogh(2004)}]{Ratanamahatana2004EverythingYK}
Chotirat~Ann Ratanamahatana and Eamonn Keogh. 2004.
\newblock Everything you know about dynamic time warping is wrong.
\newblock In \emph{Third Workshop on Mining Temporal and Sequential Data}.

\bibitem[{Rosa et~al.(2017)Rosa, Fugmann, Pinto, and Nunes}]{8037426}
Marcelo Rosa, Elmar Fugmann, Gisele Pinto, and Maria Nunes. 2017.
\newblock \href {https://doi.org/10.1109/EMBC.2017.8037426} {An anchored
  dynamic time-warping for alignment and comparison of swallowing acoustic
  signals}.
\newblock In \emph{2017 39th Annual International Conference of the IEEE
  Engineering in Medicine and Biology Society (EMBC)}, pages 2749--2752.

\bibitem[{Sakoe and Chiba(1990)}]{Sakoe:1990:DPA:108235.108244}
Hiroaki Sakoe and Seibi Chiba. 1990.
\newblock Dynamic programming algorithm optimization for spoken word
  recognition.
\newblock In Alex Waibel and Kai-Fu Lee, editors, \emph{Readings in Speech
  Recognition}, pages 159--165. Morgan Kaufmann, San Francisco.

\bibitem[{Sakurai et~al.(2007)Sakurai, Faloutsos, and Yamamuro}]{4221753}
Yasushi Sakurai, Christos Faloutsos, and Masashi Yamamuro. 2007.
\newblock \href {https://doi.org/10.1109/ICDE.2007.368963} {Stream monitoring
  under the time warping distance}.
\newblock In \emph{2007 IEEE 23rd International Conference on Data
  Engineering}, pages 1046--1055.

\bibitem[{Sart et~al.(2010)Sart, Mueen, Najjar, Keogh, and
  Niennattrakul}]{5694075}
Doruk Sart, Abdullah Mueen, Walid Najjar, Eamonn Keogh, and Vit Niennattrakul.
  2010.
\newblock \href {https://doi.org/10.1109/ICDM.2010.21} {Accelerating dynamic
  time warping subsequence search with gpus and fpgas}.
\newblock In \emph{2010 IEEE International Conference on Data Mining}, pages
  1001--1006.

\bibitem[{Schmidt et~al.(2019)Schmidt, Dietsche, Ponzetto, and
  Glava{\v{s}}}]{schmidt-etal-2019-seagle}
Fabian~David Schmidt, Markus Dietsche, Simone~Paolo Ponzetto, and Goran
  Glava{\v{s}}. 2019.
\newblock \href {https://doi.org/10.18653/v1/D19-3034} {{SEAGLE}: A platform
  for comparative evaluation of semantic encoders for information retrieval}.
\newblock In \emph{Proceedings of the 2019 Conference on Empirical Methods in
  Natural Language Processing and the 9th International Joint Conference on
  Natural Language Processing (EMNLP-IJCNLP): System Demonstrations}, pages
  199--204, Hong Kong, China. Association for Computational Linguistics.

\bibitem[{Sennrich et~al.(2016)Sennrich, Haddow, and
  Birch}]{sennrich-etal-2016-neural}
Rico Sennrich, Barry Haddow, and Alexandra Birch. 2016.
\newblock \href {https://doi.org/10.18653/v1/P16-1162} {Neural machine
  translation of rare words with subword units}.
\newblock In \emph{Proceedings of the 54th Annual Meeting of the Association
  for Computational Linguistics (Volume 1: Long Papers)}, pages 1715--1725,
  Berlin, Germany. Association for Computational Linguistics.

\bibitem[{Shieh and Keogh(2008)}]{ShiehKeogh08}
Jin Shieh and Eamonn Keogh. 2008.
\newblock {iSAX}: Indexing and mining terabyte sized time series.
\newblock In \emph{Proceedings of the 14th ACM SIGKDD International Conference
  on Knowledge Discovery and Data Mining}, KDD ’08, pages 623--631, New York,
  NY, USA. Association for Computing Machinery.

\bibitem[{Silva and Batista(2016)}]{PrunedDTW}
Diego Silva and Gustavo Batista. 2016.
\newblock \href {https://doi.org/10.1137/1.9781611974348.94} {Speeding up
  all-pairwise dynamic time warping matrix calculation}.
\newblock In \emph{Proceedings of the 2016 {SIAM} International Conference on
  Data Mining, Miami, Florida, USA, May 5-7, 2016}, pages 837--845.

\bibitem[{Snell et~al.(2017)Snell, Swersky, and
  Zemel}]{Snell2017PrototypicalNF}
Jake Snell, Kevin Swersky, and Richard~S. Zemel. 2017.
\newblock Prototypical networks for few-shot learning.
\newblock In \emph{NIPS}.

\bibitem[{Sung et~al.(2017)Sung, Yang, Zhang, Xiang, Torr, and
  Hospedales}]{Sung2017LearningTC}
Flood Sung, Yongxin Yang, Li~Zhang, Tao Xiang, Philip H.~S. Torr, and
  Timothy~M. Hospedales. 2017.
\newblock Learning to compare: Relation network for few-shot learning.
\newblock \emph{2018 IEEE/CVF Conference on Computer Vision and Pattern
  Recognition}, pages 1199--1208.

\bibitem[{Tavenard et~al.(2017)Tavenard, Faouzi, and Vandewiele}]{tslearn}
Romain Tavenard, Johann Faouzi, and Gilles Vandewiele. 2017.
\newblock tslearn: A machine learning toolkit dedicated to time-series data.
\newblock \url{https://github.com/rtavenar/tslearn}.

\bibitem[{Tibshirani et~al.(2002)Tibshirani, Hastie, Narasimhan, and
  Chu}]{10.2307/3058706}
Robert Tibshirani, Trevor Hastie, Balasubramanian Narasimhan, and Gilbert Chu.
  2002.
\newblock \href {http://www.jstor.org/stable/3058706} {Diagnosis of multiple
  cancer types by shrunken centroids of gene expression}.
\newblock \emph{Proceedings of the National Academy of Sciences of the United
  States of America}, 99(10):6567--6572.

\bibitem[{Tjong Kim~Sang and De~Meulder(2003)}]{10.3115/1119176.1119195}
Erik~F. Tjong Kim~Sang and Fien De~Meulder. 2003.
\newblock \href {https://doi.org/10.3115/1119176.1119195} {Introduction to the
  conll-2003 shared task: Language-independent named entity recognition}.
\newblock In \emph{Proceedings of the Seventh Conference on Natural Language
  Learning at HLT-NAACL 2003 - Volume 4}, CONLL '03, page 142–147, USA.
  Association for Computational Linguistics.

\bibitem[{Vanderbeck et~al.(2011)Vanderbeck, Bockhorst, and
  Oldfather}]{Vanderbeck2011AML}
Scott Vanderbeck, Joseph Bockhorst, and Chad Oldfather. 2011.
\newblock A machine learning approach to identifying sections in legal briefs.
\newblock In \emph{MAICS}.

\bibitem[{Vaswani et~al.(2017{\natexlab{a}})Vaswani, Shazeer, Parmar,
  Uszkoreit, Jones, Gomez, Kaiser, and
  Polosukhin}]{DBLP:journals/corr/VaswaniSPUJGKP17}
Ashish Vaswani, Noam Shazeer, Niki Parmar, Jakob Uszkoreit, Llion Jones,
  Aidan~N. Gomez, Lukasz Kaiser, and Illia Polosukhin. 2017{\natexlab{a}}.
\newblock \href {http://arxiv.org/abs/1706.03762} {Attention is all you need}.
\newblock \emph{CoRR}, abs/1706.03762.

\bibitem[{Vaswani et~al.(2017{\natexlab{b}})Vaswani, Shazeer, Parmar,
  Uszkoreit, Jones, Gomez, Kaiser, and Polosukhin}]{vaswani2017attention}
Ashish Vaswani, Noam Shazeer, Niki Parmar, Jakob Uszkoreit, Llion Jones,
  Aidan~N. Gomez, Lukasz Kaiser, and Illia Polosukhin. 2017{\natexlab{b}}.
\newblock \href {http://arxiv.org/abs/1706.03762} {Attention is all you need}.

\bibitem[{Vintsyuk(1968)}]{vintsyuk1968speech}
Taras~K. Vintsyuk. 1968.
\newblock Speech discrimination by dynamic programming.
\newblock \emph{Kibernetika}, 4(1):81--88.

\bibitem[{Wang and Jiang(1994)}]{wang1994complexity}
Lusheng Wang and Tao Jiang. 1994.
\newblock On the complexity of multiple sequence alignment.
\newblock \emph{Journal of computational biology}, 1(4):337--348.

\bibitem[{Xie et~al.(2016)Xie, Girshick, Doll{\'a}r, Tu, and
  He}]{Xie2016AggregatedRT}
Saining Xie, Ross~B. Girshick, Piotr Doll{\'a}r, Zhuowen Tu, and Kaiming He.
  2016.
\newblock Aggregated residual transformations for deep neural networks.
\newblock \emph{2017 IEEE Conference on Computer Vision and Pattern Recognition
  (CVPR)}, pages 5987--5995.

\bibitem[{Xu et~al.(2019)Xu, Das, and Saenko}]{Xu2019TwoStreamRC}
Huijuan Xu, Abir Das, and Kate Saenko. 2019.
\newblock Two-stream region convolutional 3d network for temporal activity
  detection.
\newblock \emph{IEEE Transactions on Pattern Analysis and Machine
  Intelligence}, 41:2319--2332.

\bibitem[{Yadav and Bethard(2018)}]{Yadav2018ASO}
Vikas Yadav and Steven Bethard. 2018.
\newblock A survey on recent advances in named entity recognition from deep
  learning models.
\newblock In \emph{COLING}.

\bibitem[{Young et~al.(2018)Young, Hazarika, Poria, and
  Cambria}]{Young2018RecentTI}
Tom Young, Devamanyu Hazarika, Soujanya Poria, and Erik Cambria. 2018.
\newblock Recent trends in deep learning based natural language processing
  [review article].
\newblock \emph{IEEE Computational Intelligence Magazine}, 13:55--75.

\bibitem[{Zhu et~al.(2017)Zhu, Klabjan, and Bless}]{zhu-etal-2017-semantic}
Xiaofeng Zhu, Diego Klabjan, and Patrick Bless. 2017.
\newblock \href {https://www.aclweb.org/anthology/I17-1095} {Semantic document
  distance measures and unsupervised document revision detection}.
\newblock In \emph{Proceedings of the Eighth International Joint Conference on
  Natural Language Processing (Volume 1: Long Papers)}, pages 947--956, Taipei,
  Taiwan. Asian Federation of Natural Language Processing.

\end{thebibliography}
